\documentclass[11pt,a4paper]{article}
\usepackage{jheppub}
\usepackage{graphicx}
\usepackage{lipsum}
\usepackage{slashed}
\usepackage{soul}
\usepackage{mathtools}
\usepackage{bm}
\usepackage[all]{hypcap}
\usepackage{tikz}
\usepackage{float}
\usepackage{tikz-feynman}
\usepackage{bbm}
\usepackage{xcolor}
\usepackage{subcaption}
\usepackage[font=small,labelfont=bf]{caption}

\usepackage{array}
\usepackage{booktabs} 

\usepackage{parskip}
\allowdisplaybreaks

\def\Z{\mathbb{Z}}

\def\Tr{\mathrm{Tr~}}

\usepackage{xspace}

\newcommand{\be}{\begin{equation}}
\newcommand{\ee}{\end{equation}}
\newcommand{\bea}{\begin{eqnarray}}
\newcommand{\eea}{\end{eqnarray}}
\newcommand{\beq}{\begin{equation}}
\newcommand{\eeq}{\end{equation}}

\newcommand{\cL}{{\cal L}}

\definecolor{revteal}{RGB}{0,110,110}

\title{Opening the Topological Portal to Dark Sectors with Colliders
}

\author[a,b]{Joe Davighi,}\emailAdd{joseph.davighi@cern.ch}
\author[c]{Admir Greljo,}\emailAdd{admir.greljo@unibas.ch}
\author[c]{Lia Sch\"oneweiß,}\emailAdd{lia.schoeneweiss@unibas.ch}
\author[d,e]{and Nud\v{z}eim Selimovi\'c\,}\emailAdd{nudzeim.selimovic@pd.infn.it}

\affiliation[a]{Theoretical Physics Department, CERN, 1211 Geneva 23, Switzerland}
\affiliation[b]{DAMTP, University of Cambridge, Wilberforce Road, Cambridge CB3 0WA}
\affiliation[c]{Department of Physics, University of Basel, Klingelbergstrasse 82, CH 4056 Basel, Switzerland}
\affiliation[d]{Dipartimento di Fisica e Astronomia, Università degli Studi di Padova, \\
Via Marzolo 8, 35131 Padova, Italy}
\affiliation[e]{Istituto Nazionale di Fisica Nucleare (INFN), Sezione di Padova, \\
Via Marzolo 8, 35131 Padova, Italy}

\abstract{We study the phenomenology of the topological portal between QCD and pseudo-Nambu--Goldstone dark matter proposed in~\cite{Davighi:2024zip}. We construct a weakly coupled ultraviolet completion in which a vector mediator gauges a baryonic current of QCD for the light quark flavours, linking it to dark pions parametrizing the coset $SU(2)_D/U(1)_D$. As a concrete anomaly-free realisation, we gauge the leptophobic combination $B_1+B_2-2B_3$. This completion allows us to calculate thermal coannihilation beyond the regime of validity of the pion effective theory, including resonant mediator effects, and to test the resulting thermal target with boosted-dijet and monojet searches at the LHC, with LEP and bottomonium data, and with flavour observables. These collider signatures follow directly from the symmetry structure required by matching onto the topological portal. At low momentum transfer, the topological operator instead controls the decay of the heavier dark pion, which is long-lived. A future high-statistics $Z$ factory, such as FCC-ee, can probe the low-mass region where hadron-collider sensitivity deteriorates, providing complementary tests of the topological portal.}

\begin{document}

\maketitle
\section{Introduction}

In the quest to understand the nature of dark matter (DM), a productive strategy is to attempt to classify the relevant portal interactions through which candidate DM particles could interact with the Standard Model (SM) degrees of freedom~\cite{Cirelli:2024ssz}. 
One natural way to structure that classification is using dimensional analysis, {\em i.e.} to classify portals by their operator dimension and hence their degree of relevance in the infrared. Indeed, if we restrict our list to portals with dimension $\leq 4$, there are only three options: the vector portal, the scalar portal, and the fermion portal. If we extend our classification to portals of higher-dimension in effective field theory (EFT), we encounter new possibilities. Notably, axion-like particle (ALP) portals start at dimension 5, and so on.

\subsection*{Lessons from $\chi$PT}

This approach to exploring DM, {\em i.e.} ordering portals by mass dimension, is rooted in the general ideas and reasoning of EFT. 
The construction of Callan, Coleman, Wess and Zumino (CCWZ) systematised the EFT expansion in $E/\Lambda$ for theories of pseudo Goldstone bosons~\cite{Coleman:1969sm,Callan1969}, by constructing all operators up to a given dimension that are invariant under the global symmetry. In the case of chiral symmetry breaking $SU(n)_L \times SU(n)_R \to SU(n)_{L+R}$, the pions package into a field $U=\exp(2i \pi^a(x)t^a/f_\pi)$ that transforms as $U\mapsto g_L U g_R^\dagger$, and the EFT expansion begins
\begin{equation} \label{eq:CCWZ}
    L \supset \frac{f_\pi^2}{4}\Tr [D_\mu U D^\mu U^\dagger] + c_1 \Tr[D_\mu U D^\mu U^\dagger]^2 + c_2 \Tr[D_\mu U D^\mu U^\dagger D_\nu U D^\nu U^\dagger] + \dots
\end{equation}
However, it was soon realised that some important physical effects were {\em not} captured by the CCWZ expansion~\cite{Wess1971}. 
To match anomalies required the inclusion of a Lagrangian operator, the Wess--Zumino--Witten (WZW) term~\cite{Wess1971,Witten1983a}, that is not invariant under the global symmetry, but shifts by a total derivative. A piece of that term reads,
\begin{equation} \label{eq:WZW_gauged}
    S_{\text{WZW}}[\pi^a, f] = \frac{N e^2}{8\pi^2 f_\pi} \int_{M_4} \pi^0 f\wedge f + \dots
\end{equation}
where $f$ is the QED field strength, $e$ the electromagnetic gauge coupling, and $M_4$ is the spacetime manifold. The action \eqref{eq:WZW_gauged} is clearly pseudo-invariant under shifting $\pi^0$. This interaction is responsible for the short lifetime of $\pi^0$ via the decay to $\gamma \gamma$.

Such pseudo-invariant interactions also have the property of being {\em topological}, meaning that they do not require a metric.
The topological action can be cast in a manifestly {\em in}variant form if one dispenses with locality;  introduce an auxiliary dimension and do the higher-dimensional integral
\begin{equation} \label{eq:WZW_5d}
    S_{\text{WZW}}[\pi^a,\, e \to 0] = \frac{-i N}{480 \pi^3} \int_{X_5} \Tr (U^\dagger dU)^5, \qquad \partial X_5 = M_4
\end{equation}
Consistency requires the coefficient $N$ be an integer~\cite{Witten1983a,Alvarez1985}. Anomaly matching further requires $N$ equals the number of colours in the underlying $SU(n_c)$ theory of quarks and gluons, which can be used to `postdict' the $\pi^0$ decay rate from~\eqref{eq:WZW_gauged}.
This manifests an important property of topological terms: being typically invariant under renormalisation group (RG) flow, they provide robust matching conditions between UV and IR. In contrast, the non-topological Wilson coefficients $c_{1,2}$ in~\eqref{eq:CCWZ} cannot be predicted from QCD with known methods. Viewed from the bottom, topological interactions thus provide unique windows onto the UV physics regardless of strong coupling regimes in between.

\subsection*{Topological portals for GeV dark matter}

Let us return to the question of dark matter coupled to the SM. In addition to the natural foliation of the space of EFT operators by operator dimension, there is a distinct, well-defined splitting of the EFT Lagrangian into a topological part and a non-topological part. As we learnt from QCD, the topological part can be important for infrared phenomena. It is therefore of interest to ask whether there exist relevant topological portals that connect DM to the SM, that might otherwise be missed in a CCWZ-style approach to classifying portals only by operator dimension. 
One should then address whether such portals give rise to viable DM models with novel phenomenology, that might suggest new strategies in our hunt for DM.

In~\cite{Davighi:2024zip} we identified one such topological portal, based on algebraic classifications of topological interactions in sigma models~\cite{DHoker:1994rdl,Davighi:2018inx,Davighi:2020vcm}. 
Suppose that DM consists of pNGBs on a vacuum manifold $M \cong G_D/H_D$ with non-trivial $H^2(M; \Z)$, such as $S^2 \cong SU(2)_D/U(1)_D$. Then the action admits a topological portal connecting the dark pions to the familiar pions of QCD. Using the same trick of extending to an extra dimension as for the WZW term~\eqref{eq:WZW_5d}, the portal action can be written
\begin{equation} \label{eq:portal_intro}
    S_{\text{portal}} = 2\pi N   \int_{X_5} \frac{1}{24 \pi^2 } \Tr(U^\dagger dU)^3 \wedge \Omega_M, 
    \qquad \partial X_5 = \Sigma_4, \qquad N \in \Z,
\end{equation}
where $\Omega_M$ is a unit-normalised differential form representing the generator of $H^2(M; \Z)$. For the particular case $M=S^2$ that was studied explicitly in~\cite{Davighi:2024zip}, it is the volume form on the sphere, $\Omega_{S^2} = \frac{1}{4\pi f_D^2}\cos(\chi_1/f_D) d\chi_1 \wedge d\chi_2$ in a particular coordinate chart. 
Here the $S^2$ radius $f_D$ is the same dark pion decay constant inferred from the kinetic term.

Expanding~\eqref{eq:portal_intro} locally, and gauging QED, the leading perturbative piece is
\begin{equation} \label{eq:2to2}
    \mathcal{L}_{\text{portal}} = \frac{eN}{16 \pi^2 f_\pi f_D^2} \epsilon^{\mu\nu\rho\sigma }\left(\pi^0 + \frac{\eta}{\sqrt{3}}\right) f_{\mu \nu} \partial_\rho \chi_1 \partial_\sigma \chi_2 + \dots
    \qquad 
\end{equation} 
for 3 flavour QCD.
The Lagrangian~\eqref{eq:2to2} provides a topological portal for the channel
\begin{equation}
    \pi^0 \gamma \leftrightarrow \chi_1 \chi_2\, ,
\end{equation}
that has not been studied before. DM can freeze-out via this $2\to 2$ channel. 
It is natural for there to be a small mass splitting between $\chi_1$ and $\chi_2$, controlled by some explicit symmetry breaking spurion parameter. In this paper, we consider a mass splitting $\Delta m_\chi \gtrsim m_\pi ~\sim \mathcal{O}(100 \text{~MeV})$, which means the $\chi_2$ particles will decay to $\chi_1$ soon after freeze-out via the topological interaction (Sec.~\ref{sec:chi2-decays}), leaving $
\chi_1$ as the stable relic DM.

More broadly, nearly degenerate dark sectors with off-diagonal interactions
provide a well-motivated setting for inelastic and co-annihilating thermal dark
matter~\cite{Griest:1990kh,DAgnolo:2018wcn,Filimonova:2022pkj}.
The topological structure of the interaction implies two features that make it well-suited to realise thermal DM with masses around the GeV range~\cite{Tucker-Smith:2001myb,Izaguirre:2015zva}: (i) it is antisymmetric in two distinct DM fields, meaning we automatically realise DM co-annihilation (the elastic channel is absent at tree-level), and (ii) it is derivative suppressed. Both facts follow inevitably from the Lagrangian being a differential form, and both enable the DM to exist in the GeV range without conflicting null results from direct or indirect detection. For the masses (and mass splittings) that we consider, up-scattering of the non-relativistic $\chi_1$ particles to $\chi_2$ off nuclei in direct detection experiments is kinematically forbidden by the $\mathcal{O}(100~ \text{MeV})$ mass splitting, see Sec.~\ref{sec:DD}. Moreover, the elastic annihilation channel $\chi_1 \chi_1 \to \text{SM}$ requires two insertions of the portal, leading to negligible emission from galaxies that might show up in indirect detection experiments. We nevertheless revisit both direct and indirect detection in \S\ref{sec:DD} and \S\ref{sec:ID} of this paper. 

These attractive features are shared by other models in which the leading DM interaction with the SM is via a differential form: for example in~\cite{Davighi:2025awm}, gauging of a dark baryon number current gives rise to a topological coupling of three dark pions to a dark photon, which can realise DM thermal freeze-out via semi-annihilation.
It is also worth noting that a QCD-like dark sector will generically feature its own WZW term~\eqref{eq:WZW_5d}; the distinctive pentavalent vertices give $3\to 2$ processes that can realise the strongly-interacting-massive-particle (SIMP) thermal history for DM~\cite{Hochberg:2014dra,Hochberg:2014kqa}. Dark matter protected by approximate global symmetries has also been explored
in composite, Higgs-portal, and vector-portal realizations, with phenomenology
that depends on how the symmetry is explicitly broken~\cite{Frigerio:2012uc,Gross:2017dan,Huitu:2018gbc}.

\subsection*{Collider probes and the need to UV complete}

A na\"ive calculation of the DM relic abundance, assuming thermal freeze-out via the contact interaction~\eqref{eq:2to2}, suggests that $\chi_1$ can constitute all the observed DM when $f_D \approx 6 m_\chi$~\cite{Davighi:2024zip}. 
For couplings of this size, the dark matter would be visible in existing $e^+e^-$ collider experiments if it is too light, being produced (alongside a $\pi^0$) from a virtual photon via the topological portal. A re-interpretation of data from BaBar~\cite{BaBar:2014zli} in the channel $e^+ e^- \to \gamma + \slashed{E}_T$ requires that $m_\chi$ be at least a few GeV~\cite{Davighi:2024zip}.\footnote{Although the exclusive $\pi^0\gamma+\slashed{E}_T$ topology considered here is
distinct from a dark-shower jet, it belongs to the broader collider program in
which hidden-sector decays partition energy between visible and invisible
states~\cite{Cohen:2015toa,Schwaller:2015gea,CMS:2021dzg}.} But for such large DM masses we run into a sharp problem of EFT validity, because pions are no longer the relevant QCD degrees of freedom. It is therefore essential to UV complete the effective topological portal Lagrangian~\eqref{eq:portal_intro} to reliably estimate the DM relic abundance in this model. 

As mentioned above, a topological action with integer-quantised coefficient must be robustly connected to a topological quantity of the UV theory, in the way that the WZW term on $G/H$ matches chiral anomalies in QCD. But the topological portal~\eqref{eq:portal_intro} cannot match any mixed anomaly between the QCD flavour and dark flavour symmetries, because there is no such mixed anomaly.
The resolution to this EFT puzzle was found in~\cite{Davighi:2024zjp}. The non-trivial 2-cycles implied by $H^2(G_D/H_D)=
\Z$ correspond to the existence of dark-sector strings charged under a generalised 1-form symmetry, and the topological portal twists this 1-form symmetry with the QCD flavour symmetry into what is known as a 2-group symmetry~\cite{Kapustin:2013uxa,Cordova:2018cvg,Benini:2018reh}.
Any valid UV completion must match this 2-group symmetry class, precisely analogous to the well-known constraints coming from anomaly matching.

Guided by this structure, an explicit UV completion for the topological portal was constructed in~\cite{Davighi:2024zjp}, albeit in a toy version, wherein an abelian gauge field couples to SM baryon number, which is matched by the topological current $j_B \propto \star \Tr (U^\dagger dU)^3$ in the IR, and to a $N_f=2$ flavour scalar QED current on the dark sector side. Strikingly, the UV completion does {\em not} require extra coloured or charged fermions (as might be expected if the portal matched a mixed anomaly) or indeed any other highly visible state, but only a weakly coupled $Z^\prime$. 

Integrating out a $Z^\prime$ with these particular couplings generates the topological portal operator, and the matching onto its coefficient must be tree-level exact because its coefficient is quantised. 

In this paper we use this construction to build a UV completion of the topological dark portal, consistent with the SM embedding above the weak scale. The $Z^\prime$ gauges the leptophobic, generation-dependent baryon-number combination $B_1+B_2-2B_3$.
Armed with a UV completion, we explore the phenomenology across a range of scales. 
For processes where the momentum transfer is large, such as scattering, we use the appropriate description featuring quarks and gluons. We use this to
revisit the relic abundance, and study collider limits from the relevant LHC $Z^\prime$ searches. With these constraints we define the region of interest for this model, for which we go on to elucidate its distinctive signatures in Belle II and FCC-ee.
For processes where the relevant momentum transfer is low, such as the decay of $\chi_2$ given small mass splitting, the topological portal Lagrangian~\eqref{eq:portal_intro} is the appropriate description and dictates the decay $\chi_2 \to \chi_1 \pi^0 \gamma$, in competition with the leptonic mode induced by kinetic mixing. The resulting lifetime is long: over most of the thermal target $\chi_2$ escapes the detector and the dark pair appears as missing energy, while at the upper edge of the viable couplings and splittings it decays at a displaced vertex in $e^+e^-$ colliders. We are thus able to give a much more robust assessment of such dramatic experimental signatures, first proposed in~\cite{Davighi:2024zip}. Compressed inelastic sectors similarly motivate searches for excited states
that decay with macroscopic lifetimes, both at flavour factories and across
the LHC long-lived-particle program~\cite{Berlin:2018jbm,Duerr:2019dmv,Alimena:2019zri}.

The structure of the rest of the paper is as follows. In Sec.~\ref{sec:topological}, we review the essential features of the symmetry matching that guides model building and present one concrete model example. In Sec.~\ref{sec:freezeout}, we perform the dark matter thermal relic abundance calculations, providing the target in the parameter space. Sec.~\ref{sec:colliders} is devoted to colliders: after determining the fate of the excited dark state (Sec.~\ref{sec:chi2-decays}), we recast the most relevant LHC, bottomonium and LEP searches for the vector boson (Sec.~\ref{sec:LHC}), discuss model-specific flavour bounds (Sec.~\ref{sec:additional}), and estimate the prospects at FCC-ee (Sec.~\ref{sec:lepton-colliders}). Finally, in Sec.~\ref{sec:conclusions} we conclude.

\section{UV completion of the topological portal}
\label{sec:topological}

Here we describe how to UV complete the IR topological portal interaction~\eqref{eq:portal_intro}. We do this by adapting the toy model set forth in~\cite{Davighi:2024zjp} to the context of an extension of the Standard Model (SM), and explain how it reproduces the topological portal proposed in~\cite{Davighi:2024zip} in the IR. We begin in \S \ref{sec:uv-general} by describing the essential features that {\em any} UV completion should have, dictated by low-energy symmetry structures of the topological coupling. Then in \S \ref{sec:UV-concrete} we set out a concrete model.

\subsection{Essential features from symmetry matching} \label{sec:uv-general}

The essential idea is simple. Our portal interaction~\eqref{eq:portal_intro} couples the topologically conserved baryon current of QCD to the dark topological two-form current that generates a one-form symmetry. It can be rewritten, locally, in the suggestive form
\begin{equation}
    S_{\mathrm{portal}} = -2 \pi N \int_{M_4} \star J_B \wedge A_D, 
\end{equation}
where 
\begin{equation} \label{eq:AD}
    A_D = \frac{1}{8\pi f_D^2}(\chi_1 d\chi_2 - \chi_2 d\chi_1) + \mathcal{O}(\chi^3)
\end{equation}
is a locally-defined potential 1-form for the integral 2-form, $dA_D=\Omega_{S^2}$. (Note that $2\pi A_D$ is the properly-normalised connection.) 
One would expect to obtain such a term by coupling the conserved baryon number current $J_B$ of QCD to a $U(1)_X$ gauge field $X_\mu$ whose equation of motion sets $X \propto A_D + ...$, and integrating out that gauge field. 

We know how to interpolate baryon current above the QCD scale, and so we know how the gauge field $X_\mu$ should couple to the light quarks in any UV completion. By `light', we refer to the $u, d$, and $s$ quarks that determine the chiral Lagrangian. For a clean embedding in the SM, we should at least extend this to charm also by $SU(2)_L$ invariance, but we can consider third generation baryon number separately because the top and bottom quarks are integrated out above the QCD scale.

On the dark matter side, the low-energy theory (and its symmetries) is not enough to pin down the origin of the dark pNGBs $\chi_{1,2}$. 
For example, the pNGBs might emerge as the angular modes of weakly coupled scalars (as we set out concretely in \S \ref{sec:UV-concrete}), or they might emerge from some strongly coupled dark sector (more analogous to the pions of QCD).
What we do know, unambiguously, is the induced connection $2\pi A_D$ on the pNGB manifold, and from there we can reconstruct a St\"uckelberg action for the gauge field coupling to the Goldstones.

Putting things together, we know that, above the QCD scale, our effective Lagrangian must feature couplings
\begin{align}
    \cL \supset \frac{F_X^2}{2}\left(\partial_\mu\vartheta + 2\pi A_{D,\mu} + x_\phi g_X X_\mu \right)^2 + x_q g_X X_\mu J_q^\mu, \qquad  J_{q}^\mu = \sum_{q=u,d,s,c} \overline{q}\gamma^\mu q \, .
\end{align}
for now written in an arbitrary gauge where we retain the compact St\"uckelberg field $\vartheta$ that gets eaten. The `decay constant' $F_X$ parametrises the normalisation of that $\vartheta$ direction, which can in principle differ from $f_D$ depending on the UV completion; the factor $F_X$ will nevertheless drop out upon matching to the portal.
Here $g_X$ is the $U(1)_X$ gauge coupling and ($x_q$, $x_\phi$) are integer charges. In fact, to see the correct quantisation of the portal operator emerge, it is convenient to choose $x_\phi=1$ for now, and consider generalisation to other $x_\phi$ after.

Going to unitary gauge ($\vartheta=0$),
the leading order EOM for the heavy gauge field, neglecting in particular its kinetic term, is 
\begin{equation}
    X^\mu =\frac{1}{g_X}\left( - 2\pi A_D^\mu
    -\frac{x_q}{F_X^2}J_q^\mu\right)\, .
\end{equation}
After integrating out $X^\mu$, we get the dimension-6 EFT Lagrangian
\begin{equation}  \label{eq:Ldim6}
    \cL_{\rm EFT}\supset
    -2\pi x_q A_{D,\mu} \, J_q^\mu
    -\frac{x_q^2}{2F_X^2}J_{q,\mu}J_q^\mu .
\end{equation}
We obtain operators with $\chi_1 \partial_\mu \chi_2$ coupled to the quark current (which will match to the portal), as well as four-quark operators.

At low energies, the integer-normalised quark current matches onto the topologically conserved baryon number current of low-energy QCD,
\begin{equation}
    J_q^\mu = n_c J_B^\mu, \qquad J_B = \star \frac{1}{24\pi^2} \Tr (U^\dagger dU)^3\, .
\end{equation}
The first term in~\eqref{eq:Ldim6} thus matches onto the topological portal~\eqref{eq:portal_intro} after flowing through the QCD phase transition,
with the identification of its integral coefficient:
\begin{equation} \label{eq:N}
    N = n_c x_q\in \Z\, .
\end{equation}
If we restore non-minimal values for the dark charge $x_\phi$, it crucially does {\em not} appear in the normalisation of the topological term. This is a little subtle; rescaling $x_\phi$ requires us to pass to a quotient target space $S^2/\Z_{|x_\phi|}$ and we must rewrite things in terms of a correctly normalised connection thereon. Sure enough, the factors of $x_\phi$ drop out and the coefficient is always given by~\eqref{eq:N}~\cite{Davighi:2024zjp}.

The rigidity of this EFT coefficient can be understood in terms of the topological portal matching a 2-group generalised symmetry structure of the UV theory~\cite{Davighi:2024zjp}. (A similar discussion of how this works, but in a phenomenological context, was given in~\cite{Davighi:2025awm}.) This rigidity means there are no other higher-loop contributions to the matching onto~\eqref{eq:portal_intro}; the matching is, unusually, `tree-level exact'.

This completes our discussion of the universal features of `UV' completions of the topological portal, where for now `UV' means we have ascended above $\Lambda_{\text{QCD}}$.
But, of course, what we have described is not yet a fundamental theory. There are two key aspects that should be specified to embed this EFT in a renormalisable model, but these involve choices that are not fixed by symmetries alone:
\begin{enumerate}
    \item Renormalisable origin for the dark pions. This could be a weakly coupled completion in which the pNGBs are embedded as angular modes of two complex scalar fields (\S \ref{sec:UV-concrete}), or a strongly coupled completion involving dark quarks.\footnote{
    A possible starting point for such a strongly-coupled dark sector completion is the following. Consider a pair of LH Weyl fermions $\psi_i^a$ charged in the fundamental representation of an $SO(n_c)$
    dark sector gauge group, where $a=1,\dots n_c$ is the gauge index.
    This has $SU(2)$ global symmetry, $\psi^a_i \mapsto U_i^j \psi^a_j$. 
    A gauge-invariant chiral condensate is expected to form in the channel $\langle \psi^a_i \psi^a_j \rangle \sim \Lambda^{3} \delta_{ij}$, breaking $SU(2) \to SO(2)$ and delivering our desired pair of Goldstones $\chi_i$ on $S^2$.
    However, a difficulty emerges in trying to couple the abelian gauge field $X$ to the Goldstone current $A_D$ needed to match onto the topological portal. That would require gauging a $U(1)_X$ symmetry under which $\psi_{1,2}^a$ have equal charge, which makes $U(1)_X \times SO(n_c)$ anomalous. So this framework requires additional chiral fermions to be made consistent. We do not explore this option further in the present work, being content with the weakly-coupled completion in \S \ref{sec:UV-concrete}.
    }
    \item Anomaly-free embedding of baryon number. Additional couplings of the $X_\mu$ gauge field to fermions must be present in order to render the gauge symmetry $G_{\text{SM}} \times U(1)_X$, as realised above the weak scale, anomaly-free. The most obvious embeddings are vector-like combinations of baryon and lepton numbers, such as $B-L$, $B-3L_3$, $B_1+B_2-2B_3$, and so on, with right-handed neutrinos or other additional chiral matter included where required. Gauged baryon currents and leptophobic quark portals have been studied in a variety of ultraviolet completions; anomaly cancellation commonly correlates the vector mediator with additional fermionic matter~\cite{FileviezPerez:2010gw,Dobrescu:2014fca,Ismail:2016tod,Caron:2018yzp}.
\end{enumerate}
In the next Section, we set out a fully complete concrete model that has a number of desirable phenomenological features. This involves a weakly coupled completion of the dark pNGBs, and gauging the anomaly-free $B_1+B_2-2B_3$ current.

\subsection{A concrete model: $B_1+B_2-2 B_3$} \label{sec:UV-concrete}

The scenario constructed in this Section, which is built on the essential features outlined in \S \ref{sec:uv-general}, can be regarded as a proof of existence of a UV completion that reproduces our topological portal in the IR.

The symmetry group and field content are summarized in Tab.~\ref{tab:fields}. We gauge the generation-dependent baryon-number combination $B_1+B_2-2B_3$:\footnote{We introduce a slight abuse of notation when referring to the UV model, which is common in the model-building literature; the gauged current is technically $n_c (B_1+B_2-2B_3)$ such that all quarks have integer charges. That is, in this work, we use the primitive convention for the charge normalization $x_q=1$, which differs from the usual baryon convention $x_q=1/3$.} 
first- and second-generation quarks carry charge $x_q$, third-generation quarks carry charge $-2x_q$, and all SM leptons are neutral. 
This choice is leptophobic and anomaly-free without introducing additional fermions. The dark sector comprises two complex scalars $\phi_{1,2}$, each coupled with the same charge $x_\phi$ to the abelian gauge field $X_\mu$ that mediates between the SM and the dark sector.

\begin{table}[t]
  \centering
  \small
  \renewcommand{\arraystretch}{1.18}
  \begin{tabular}{@{}>{\raggedright\arraybackslash}p{3.3cm}>{\raggedright\arraybackslash}p{4.5cm}c|cc@{}}
    \toprule
    Fields & $SU(3)_c\times SU(2)_L\times U(1)_Y$ & $U(1)_X$ & $SU(2)_D$ & $P_D$ \\
    \midrule
    $q^a_{L}, u^a_{R}, d^a_{R}$ ($a=1,2$)
      & usual quark irreps & $x_q$ & $\mathbf{1}$ & $+$ \\
   $\Phi=(\phi_1,\phi_2)^T$
      & $(\mathbf{1},\mathbf{1})_0$ & $x_\phi$ & $\mathbf{2}$ & $(-,+)$ \\
      \hline
    $q^3_{L}, u^3_{R}, d^3_{R}$
      & usual quark irreps & $-2x_q$ & $\mathbf{1}$ & $+$ \\
    $U_{L,R}$ or $D_{L,R}$
      & $(\mathbf{3},\mathbf{1})_{2/3}$, $(\mathbf{3},\mathbf{1})_{-1/3}$ & $x_q$ & $\mathbf{1}$ & $+$ \\
    \bottomrule
  \end{tabular}
  \caption{Symmetries and field content of a minimal completion of the topological portal, based on gauging $B_1+B_2-2B_3$. Here, $U(1)_X$ is gauged while $SU(2)_D$ is an approximate global symmetry: its column entries pertain to the symmetric dark-scalar sector limit (which is explicitly broken by small couplings), whereas exact parity $P_D$ is imposed on the complete Lagrangian ensuring DM stability. The SM Higgs and leptons are charged only under the SM gauge group, in the usual way. The index $a=1,2$ labels the first two SM generations; only one Dirac vector-like quark $U_{L/R}$ or $D_{L/R}$ is required to generate the CKM mixing. See \S\ref{sec:UV-concrete} for details. }
  \label{tab:fields}
\end{table}

It is useful to separate charge conventions from physical couplings.  We define
\begin{equation}
  g_q\equiv g_Xx_q,\qquad Q_\chi\equiv x_\phi,
  \qquad r_\chi\equiv\frac{x_\phi}{x_q},\qquad
  g_\chi\equiv g_XQ_\chi.
  \label{eq:coupling-convention}
\end{equation}
In Sec.~\ref{sec:freezeout}, we adopt the normalization $x_q=1$, such that $g_q=g_X$, and the dark charge $Q_\chi$ is precisely $x_\phi$.

The dark scalars are assembled into a doublet $\Phi=(\phi_1,\phi_2)^T$ of $G_D=SU(2)_D$, which is an approximate symmetry of their scalar potential. We also impose an exact ${\mathbb Z}_2$ parity
\begin{equation}
  P_D:\qquad \phi_1\longmapsto-\phi_1,
  \qquad \phi_2\longmapsto\phi_2,
  \label{eq:dark-parity}
\end{equation}
under which all other fields are even. This discrete symmetry of the potential is imposed on the complete Lagrangian and will ensure the stability of the lightest pNGB, which serves as DM. 

We emphasise that the set of field representations recorded in the first two rows of Tab.~\ref{tab:fields} is a robust assignment for any weakly-coupled completion of the topological portal;  the charges written in the remaining rows are, on the other hand, specific to this particular UV completion. In the renormalizable flavour completion adopted below, a single heavy vector-like quark $U_{L,R}$ generates both third-to-first- and third-to-second-generation mixing.

The dynamics are governed by the gauge-invariant Lagrangian, 
\begin{equation} \label{eq:LUV}
    \cL = \cL_{\text{SM}} - \frac{1}{4} X_{\mu\nu}X^{\mu\nu} 
    + g_X J_{X,\text{SM}}^\mu X_\mu
    + \sum_{i=1}^2 |D_\mu \phi_i|^2 - V\left(\phi_i^\ast \phi_j\right) + \frac{\epsilon_Y}{2} X_{\mu\nu}B^{\mu\nu} + \lambda_{H\phi}^{ij}(H^\dagger H) (\phi_i^\ast \phi_j)
\end{equation}
where $X_{\mu\nu}=\partial_\mu X_\nu-\partial_\nu X_\mu$. 
The covariant derivative acts on the doublet of complex scalars as
    $D_\mu \phi_i = (\partial_\mu - ix_\phi g_X X_\mu) \phi_i$,
and the gauge current coupling to the SM fields is
\begin{equation}
    J_{X,\text{SM}}^\mu = x_q \sum_{i=1}^2 (\overline{q}_i \gamma^\mu q_i + \overline{u}_i \gamma^\mu u_i + \overline{d}_i \gamma^\mu d_i)
    -2x_q (\overline{q}_3 \gamma^\mu q_3 + \overline{u}_3 \gamma^\mu u_3 + \overline{d}_3 \gamma^\mu d_3)\, ,
\end{equation}
corresponding to gauging $B_1+B_2-2B_3$, as reflected in Tab.~\ref{tab:fields}. 
The $[SU(3)_c]^2U(1)_X$, $[U(1)_X]^3$, gravitational, and $U(1)_Y[U(1)_X]^2$ anomalies vanish generation by generation because the quark charges are vector-like. The remaining $[SU(2)_L]^2U(1)_X$ and $[U(1)_Y]^2U(1)_X$ anomalies are proportional to the sum of the generation weights, $1+1-2=0$, and therefore cancel among the three generations. We discuss the final three terms in~\eqref{eq:LUV}, together with the generation of the full CKM matrix, in the remainder of this Section.

As we will demonstrate in \S \ref{sec:colliders}, the main constraints will arise from the jet-based collider searches, which are unavoidable predictions of the mechanism common to all models. Complementary to this, the specific UV realisation predicts the flavour-changing neutral currents in the up-sector, in particular, the rare top decays $t\to cX(\to jj)$ and $D-\bar D$ mixing. Before turning to phenomenology, we first review how the renormalisable dark sector couplings introduced in~\eqref{eq:LUV} reproduce the pNGB dark matter, coupled via the connection $2\pi A_D$ in~\eqref{eq:AD}, at low energy.

\subsubsection*{Dark symmetry breaking transition}

The potential $V(\phi_i)$ for the two complex scalars $\phi_i$ will be such that these scalars acquire a non-zero vacuum expectation value (vev) to trigger spontaneous symmetry breaking, giving rise to the dark matter particles $\chi_i$ as the Goldstone modes, as follows.

Firstly, note that if there were no gauge fields coupled to $\phi_i$, the theory would enjoy global $O(4)$ symmetry, as can be seen by repackaging the real and imaginary parts of $\phi_{1,2}$ into a 4-component real vector. 
The gauging of $U(1)_X$ that acts universally on $\phi_i$ reduces this $O(4)$ to an $SU(2)_D$ global symmetry acting as $\phi_i \mapsto U_i^j \phi_j$ for $U \in SU(2)_D$. Let us package $\phi_i$ into a complex doublet field $\phi:=(\phi_1,\phi_2)^T$.
In the limit of exact $SU(2)_D$ global symmetry, the potential would be\footnote{In \S \ref{sec:mass} we will generalise this to enforcing only gauge invariance and the exact discrete symmetry $P_D$; the resulting $SU(2)_D$-breaking terms are actually needed to generate mass for the DM pions.}
\begin{equation}\label{eq:exactSP}
    V_0(\phi) = \lambda_D \left( \phi^\dagger \phi - \frac{v_D^2}{2}\right)^2,
    \qquad
    \langle\phi\rangle=\frac{1}{\sqrt2}\begin{pmatrix}0\\v_D\end{pmatrix},
    \qquad f_D=\frac{v_D}{2}.
\end{equation}
What is the resulting pNGB manifold?
In the limit $g_X \to 0$, the vev would break $O(4) \to O(3)$. 
Restoring the gauging, the unbroken subgroup is $O(3) \cap SU(2)_D \cong U(1)_D$, and the
vacuum manifold is reduced to $SU(2)_D/U(1)_D\cong S^2$. 
The third would-be Goldstone becomes the longitudinal component of the now-massive $X$ gauge field. This structure exactly mirrors the SM electroweak theory with $SU(2)_L$ gauge coupling turned off.

In~\cite[\S 4.3.2]{Davighi:2024zjp}, a system of global homogeneous coordinates on $S^2 \cong \mathbb{C} P^1$ was used to perform a globally-valid EFT matching, topologically speaking. Here, we are interested primarily in the leading order local interactions, so we can stick to the more pedestrian `local matching' set out in~\cite[\S 4.3.1]{Davighi:2024zjp}.

Without loss of generality, we take the unbroken $U(1)_D$ direction to be generated by $\sigma^3/2$ in $\mathfrak{su}(2)_D$. That is, the $U(1)_D$ charge is identified as the combination
\begin{equation}
  Q_{D}=T_{SU(2)_D}^3+\frac{X}{2x_\phi},
  \label{eq:unbroken-dark-generator}
\end{equation}
for which $Q_D(\phi_2)=0$ and $Q_D(\phi_1)=1$. In unitary gauge, the complex doublet $\phi:=(\phi_1,\phi_2)^T$ is expanded as
\begin{equation} \label{eq:phi}
    \Phi =  e^{\frac{i}{v_D} (\chi_1 \sigma^1+\chi_2\sigma^2)}
    \frac{1}{\sqrt2}\begin{pmatrix}
    0 \\
    v_D + \rho
    \end{pmatrix} \, .
\end{equation}
where recall $f_D = v_D/2$ relates the pNGB decay constant to the vev. Importantly, dark-matter parity defined in \eqref{eq:dark-parity} is not spontaneously broken by this choice, $P_D \subset U(1)_D$. Otherwise, $\langle \phi_2 \rangle$ insertions might have destabilized dark matter. The fields $\chi_1$ and $\chi_2$ are the Goldstone modes, corresponding to the broken generators $\sigma^{1,2}$, while $\rho$ is the radial mode. At leading order in $v_D^{-1}$,
$\phi_1=(\chi_2+i\chi_1)/\sqrt2+\mathcal O(\chi^2/v_D)$ and
$\phi_2=(v_D+\rho)/\sqrt2+\mathcal O(\chi^2/v_D)$; hence both $\chi_1$ and $\chi_2$ are odd under $P_D$. Note that the action of $\sigma^3$ on the vev is equivalent to a $U(1)_X$ gauge transformation; hence, that generator is unbroken. We have chosen a gauge in which the would-be third Goldstone $\chi_3$ is eaten by the $X$ gauge field. We neglect the radial mode $\rho$, assuming it is heavy and integrated out, and focus on the interactions of the dark Goldstones $\chi_i$.

Substituting the expansion
~\eqref{eq:phi} back into the kinetic term for $\phi_i$ in~\eqref{eq:LUV}, we can infer the coupling of the gauge field to the Goldstones. We have $\phi^\dagger \partial_\mu \phi = -\frac{i}{2}(\chi_1 \partial_\mu \chi_2 - \chi_2 \partial_\mu \chi_1)$.
Keeping the quadratic terms and the leading interaction with the gauge field, we get\footnote{The leading interactions in $f^{-1}_D$ come specifically from $|D_{\mu} \phi_1|^2$, after substitution $\phi_1=(\chi_2+i\chi_1)/\sqrt 2$. The fact that both Goldstones end up embedded in $\phi_1$ allows us to deduce they both end up charged under the discrete $P_D \cong \Z_2$ parity symmetry that remains, which stabilises the pNGB dark matter particles. } 
\begin{equation}
    \cL \supset \sum_{i=1}^2 |D_\mu \phi_i|^2 = \frac{1}{2} \sum_i (\partial_\mu \chi_i)^2 + 2 x_\phi^2 g_X^2 f_D^2 X_\mu X^\mu + x_\phi g_X X_\mu (\chi_1 \partial^\mu \chi_2 - \chi_2 \partial^\mu \chi_1),
\end{equation}
matching precisely onto the relevant term in the effective St\"uckelberg Lagrangian we wrote down using only symmetries in \S \ref{sec:uv-general}, with canonically normalised fields, provided we identify the scales $F_X = 2f_D = v_D$. The vector mass in the one-condensate model is therefore
\begin{equation}
  m_X=2|g_Xx_\phi|f_D.
  \label{eq:single-condensate-mass}
\end{equation}
For primitive integer charges, the condensate leaves a discrete gauge subgroup $\Z_{|x_\phi|}^{X}$.  In the minimal flavour completion $x_q=1$ and $x_\phi=3$, so this subgroup is $\Z_3^X$, up to possible identifications with the centres of the SM gauge factors.  It is distinct from $P_D$ and does not stabilize $\chi_1$; stability follows from the independently imposed parity in Eq.~\eqref{eq:dark-parity}.

\subsubsection*{Generating the dark matter mass} 
\label{sec:mass}

The two Goldstone bosons $\chi_{1,2}$ must acquire mass, and moreover a small splitting is desirable for dark matter phenomenology.
Near the $SU(2)_D$-symmetric limit \eqref{eq:exactSP}, the relevant parity-preserving dark scalar potential can be organized as
\begin{align}
  V_D={}&V_0(\phi)
  +\mu_\chi^2|\phi_1|^2
  +\left[\frac{\kappa_D}{2}(\phi_1^\dagger\phi_2)^2+{\rm h.c.}\right]
  +\delta V_{\rm even},
  \label{eq:potential-proposed}
\end{align}
Here $\delta V_{\rm even}$ denotes further $P_D$-even and $U(1)_D$-preserving deformations built from $|\phi_1|^2$ and $|\phi_2|^2$.  Terms with an odd number of $\phi_1$ fields, such as $\phi_1^\dagger\phi_2$, $(H^\dagger H)\phi_1^\dagger\phi_2$, and the flavour coupling $\overline U_L\phi_1t_R$, are forbidden by $P_D$. This, again, is crucial for dark matter stability.

The Higgs portal separates into an $SU(2)_D$-invariant coupling and the breaking spurion
$\Delta\lambda_H\equiv\lambda_{H1}-\lambda_{H2}$.  After electroweak symmetry breaking and minimization in the $\phi_2$ direction, the latter shifts the common pNGB mass by
\begin{equation}
  \delta m_\chi^2=\frac{\Delta\lambda_H}{2}v_{\rm EW}^2,
  \label{eq:higgs-portal-mass}
\end{equation}
whereas $\lambda_{H2}$ shifts the dark and electroweak vacuum conditions and induces radial mixing $\delta m_{h\rho}^2=\lambda_{H2}v_{\rm EW}v_D$.  Thus a light pNGB requires $|\Delta\lambda_H|\lesssim2m_{\chi_1}^2/v_{\rm EW}^2$ unless cancellations are accepted.  If $\Delta\lambda_H$ is generated at a generic one-loop size, this reproduces the useful estimate $m_{\chi_1}\sim v_{\rm EW}/(4\pi)$ up to couplings and factors of order unity. 

The spurion $\mu_\chi^2 \ll \lambda_D v^2_D$ represents a small explicit breaking of $SU(2)_D$ down to $U(1)_D$. It can therefore generate a common mass for $\chi_1$ and $\chi_2$, but cannot split them. Additional $SU(2)_D$-breaking contributions including $\delta V_{\rm even}$, the Higgs portal, and the vector-like fermion interactions also preserve $U(1)_D$.  There is only one term in the renormalizable Lagrangian that can break $U(1)_D$. In particular, the $\kappa_D$ term carries two units of $U(1)_D$ charge and breaks this continuous symmetry to its $\alpha=\pi$ subgroup, precisely $P_D$.

After a phase redefinition that makes $\kappa_D$ real, which is possible because all other displayed terms are invariant under a relative rephasing of $\phi_1$ and $\phi_2$, expansion about the parity-preserving vacuum gives
\begin{equation}
  m_{\chi_1}^2=\mu_\chi^2-\frac{\kappa_Dv_D^2}{2},
  \qquad
  m_{\chi_2}^2=\mu_\chi^2+\frac{\kappa_Dv_D^2}{2},
  \qquad
  m_{\chi_2}^2-m_{\chi_1}^2=\kappa_Dv_D^2,
  \label{eq:dark-masses}
\end{equation}
where $\mu_\chi^2$ is understood to include the parity-even common-mass contributions from $\delta V_{\rm even}$, electroweak symmetry breaking, and threshold corrections. 
The displayed vacuum requires both eigenvalues to be positive; in the truncated potential this implies $\mu_\chi^2>|\kappa_D|v_D^2/2$, in addition to the usual boundedness conditions on the complete quartic potential. A phenomenologically interesting region of parameter space has a small relative splitting $\kappa_D v^2_D \ll \mu^2_\chi$. That is, the explicit breaking can be thought of in two stages:
\begin{equation}
    SU(2)_D \xrightarrow{\langle \mu_\chi \rangle,\, \dots} U(1)_D \xrightarrow{\langle \kappa_D \rangle} P_D \simeq \Z_2,
\end{equation}
leaving only a remnant discrete symmetry that stabilizes the DM particle $\chi_1$.

For $\kappa_D>0$, $\chi_1$ is the lighter parity-odd state and is exactly stable as long as $P_D$ is exact and $\langle\phi_1\rangle=0$.  The heavier odd state may decay as $\chi_2\to\chi_1 X^*\to\chi_1\,+$ parity-even SM states. 
At small momentum release, the latter must be matched onto hadrons; the characteristic topological channel is $\chi_2\to\chi_1\pi^0\gamma$, while kinetic mixing can additionally open $\chi_2\to\chi_1\ell^+\ell^-$ (both are analysed in Sec.~\ref{sec:chi2-decays}).
The limit $\kappa_D\to0$ restores $U(1)_D$, so a small splitting is technically natural. Notice that the symmetry guarantees the splitting to be proportional to $\kappa_D$; $SU(2)_D$-breaking but $U(1)_D$-preserving thresholds renormalize only the common mass when $\kappa_D=0$. From here on, we can treat the dark matter $\chi_1$ mass ($m_\chi$) and the mass splitting ($\Delta m_\chi$) as phenomenological parameters in the dark pion EFT.

\subsubsection*{Generating the full CKM matrix}
\label{sec:CKM}

Because $H$ is neutral under $U(1)_X$, renormalizable SM Yukawa couplings
connect only generations with equal $X$ charge.  The resulting $2+1$ block
structure would contain no mixing between the third and first two families. A minimal renormalizable completion that
treats the two Yukawa sectors in parallel introduces the two vector-like
weak singlets,
\begin{align}
  -\mathcal L_{\rm flav}\supset{}&
  M_U\overline U_LU_R
  +\lambda^u_a\overline Q_{La}\widetilde H U_R
  +y_U\overline U_L\phi_2t_R
  \nonumber\\
  &+M_D\overline D_LD_R
  +\lambda^d_a\overline Q_{La}H D_R
  +y_D\overline D_L\phi_2b_R+{\rm h.c.},
  \qquad a=1,2.
  \label{eq:vlq-completion}
\end{align}
The minimal model needs only one of these fields, either up-type partner $U_{L/R} \sim (\mathbf{3},\mathbf{1})_{2/3}$ or down-type partner $D_{L/R}\sim (\mathbf{3},\mathbf{1})_{-1/3}$. For generality, we keep both fields present in the theory.
Gauge invariance of the last interaction in each line fixes
\begin{equation}
  x_\phi=3x_q
  \label{eq:minimal-charge-relation}
\end{equation}
in the one-condensate model.  Both interactions select the parity-even field
$\phi_2$ and hence preserve $P_D$ and the continuous $U(1)_D$.  A single family of $U$ or $D$ is sufficient: the two components of $\lambda^u_a$ and
$\lambda^d_a$ independently generate mixing with the first and second families.

For $M_{U,D}$ above the symmetry-breaking scales, integrating out the heavy
fermions gives
\begin{equation}
  \mathcal L_{\rm eff}\supset\phi_2\left(
  \frac{\lambda^u_ay_U}{M_U}\overline Q_{La}\widetilde Ht_R
  +\frac{\lambda^d_ay_D}{M_D}\overline Q_{La}Hb_R\right)+{\rm h.c.}
  \label{eq:flavour-effective-operator}
\end{equation}
After both symmetries are broken, these operators generate the mass-matrix
entries
\begin{equation}
  (M_u)_{a3}=\delta^u_a
  =-\frac{\lambda^u_ay_Uv_{\rm EW}v_D}{2M_U},
  \qquad
  (M_d)_{a3}=\delta^d_a
  =-\frac{\lambda^d_ay_Dv_{\rm EW}v_D}{2M_D}.
  \label{eq:quark-mass-mixings}
\end{equation}
At leading order the corresponding left-handed angles scale as
$\theta^{u,L}_{a3}\simeq\delta^u_a/m_t$ and
$\theta^{d,L}_{a3}\simeq\delta^d_a/m_b$.  The unrestricted relative rotation
of the light $2\times2$ blocks supplies the Cabibbo angle.

The distribution of CKM mixing between the two sectors is most transparently
described without choosing a weak basis.  Write
$u_L^{\rm gauge}=U_u u_L$ and $d_L^{\rm gauge}=U_d d_L$, so that
$V_{\rm CKM}=U_u^\dagger U_d$, and define the unit vectors
\begin{equation}
  a\equiv U_u^\dagger e_3,
  \qquad b\equiv U_d^\dagger e_3,
  \qquad a=V_{\rm CKM}b,
  \qquad e_3=(0,0,1)^T.
  \label{eq:flavour-alignment-vectors}
\end{equation}
Since $\mathrm{diag}(1,1,-2)=\mathbbm 1-3e_3e_3^\dagger$, the left-handed
$X$ currents in the two mass bases are the rank-one projectors
\begin{equation}
  \Gamma_{u_L}=g_q\bigl(\mathbbm 1-3aa^\dagger\bigr),
  \qquad
  \Gamma_{d_L}=g_q\bigl(\mathbbm 1-3bb^\dagger\bigr),
  \qquad g_q\equiv g_Xx_q.
  \label{eq:mass-basis-flavour-currents}
\end{equation}
Equation~\eqref{eq:mass-basis-flavour-currents} contains the complete
tree-level flavour information relevant below.  The frequently used
down-aligned limit, $b=e_3$ and $a_i=V_{ib}$, gives
$(\Gamma_{u_L})_{ij}=g_q\delta_{ij}-3g_qV_{ib}V_{jb}^*$ and a diagonal down
current.  It is a phenomenologically convenient benchmark, predicted when only the field $U_{L/R}$ is present in the theory.  More generally, reducing the $u$--$c$ current necessarily turns on
$d$--$s$, $d$--$b$, or $s$--$b$ currents. We quantify this alignment trade-off in
Sec.~\ref{sec:additional}. Right-handed flavour-changing
currents are parametrically suppressed by additional mass ratios for the triangular textures in
Eq.~\eqref{eq:quark-mass-mixings}.  

The minimal relation $r_\chi=x_\phi/x_q=3$ may be too restrictive when a large invisible vector width is desired.  The flavour and dark Higgs sectors can be separated by replacing $\phi_2$ in Eq.~\eqref{eq:vlq-completion} with an $SU(2)_D$-singlet scalar $S$ of charge $3x_q$ and vacuum expectation value $v_S/\sqrt2$.  The dark doublet may then have an independent charge $x_\phi$, and
\begin{equation}
  m_X^2=g_X^2\left(x_\phi^2v_D^2+9x_q^2v_S^2\right).
  \label{eq:two-condensate-mass}
\end{equation}
This extension is not obtained by the replacement $x_\phi\to3x_q$ in the one-condensate formula: it contains an additional radial mode and, unless lifted by scalar interactions, an additional physical phase.  The phenomenological scans in Sec.~\ref{sec:freezeout} with $x_q=1$ and $Q_\chi\ne3$ are to be interpreted in this separated-condensate completion.  In particular, for the benchmark $Q_\chi=x_\phi=1$, the renormalizable interaction
\begin{equation}
  V(\Phi,S)\supset\left[\lambda_{S\Phi}S^\dagger\phi_2^3+{\rm h.c.}\right]
  \label{eq:phase-locking}
\end{equation}
is gauge invariant, preserves $P_D$, and removes the otherwise massless relative phase.  Other charge ratios require an appropriate gauge-invariant phase-locking operator, which may be nonrenormalizable.  With both condensates the residual subgroup inside $U(1)_X$ is $\Z_{\gcd(|x_\phi|,3|x_q|)}^X$.

\subsubsection*{Other portals}

The exact symmetries allow both hypercharge--$X$ kinetic mixing and the diagonal Higgs portal; neither is consistently eliminated merely by omitting it at one scale.  For small $\epsilon_Y$, the one-loop running in the unbroken theory is~\cite{Holdom:1985ag}
\begin{equation}
  \mu\frac{d\epsilon_Y}{d\mu}
  =\frac{g_Yg_X}{16\pi^2}
  \left[
    \frac{2}{3}\operatorname{Tr}_{\rm Weyl}(YX)
    +\frac{1}{3}\operatorname{Tr}_{\rm scalar}(YX)
  \right].
  \label{eq:kinetic-mixing-rge}
\end{equation}
The dark scalars have $Y=0$, while the SM-quark trace, including colour and weak multiplicities, is
\begin{equation}
  \operatorname{Tr}_{\rm SM}(YX)
  =2(x_1+x_2+x_3)=0.
  \label{eq:sm-kinetic-trace}
\end{equation}
Thus the logarithmic SM contribution cancels when all quark flavours are
active.  The vector-like quarks do not separately share this cancellation.
Their hypercharges give
\begin{equation}
  \Delta\operatorname{Tr}(YX)\big|_U=4x_q,
  \qquad
  \Delta\operatorname{Tr}(YX)\big|_D=-2x_q,
  \qquad
  \left.\mu\frac{d\epsilon_Y}{d\mu}\right|_{U+D}
  =\frac{g_Yg_Xx_q}{12\pi^2}\,.
  \label{eq:vlq-kinetic-mixing}
\end{equation}
above both thresholds.  More generally, imposing $\epsilon_Y(\Lambda)=0$ at
$\Lambda>M_U,M_D$ gives the leading-log estimate
\begin{equation}
  \epsilon_Y\sim
  \frac{g_Yg_Xx_q}{6\pi^2}\ln\!\frac{\Lambda}{M_U}
  -\frac{g_Yg_Xx_q}{12\pi^2}\ln\!\frac{\Lambda}{M_D},
  \label{eq:kinetic-mixing-estimate}
\end{equation}
up to the appropriate threshold matching.
Below electroweak symmetry breaking, unequal quark thresholds produce further finite photon--$X$ mixing even though the ultraviolet SM trace vanishes.  For $m_X\ll m_Z$, writing $\epsilon_\gamma\simeq c_W\epsilon_Y$, canonical normalization induces the interaction
$\epsilon_\gamma e X_\mu J_{\rm em}^\mu$, including couplings to charged leptons~\cite{DEramo:2016gos}.

The original $B_1+B_2-2B_3$ coupling dominates the coupling of $X$ to a fermion $f$ provided
\begin{equation}
  |\epsilon_Y|g_Y|Y_f|\ll g_X|x_f|,
  \label{eq:kinetic-mixing-hierarchy}
\end{equation}
or, parametrically, $|\epsilon_Y|\ll g_X|x_q|/g_Y$ for the light quarks.  Exact leptophobia is not radiatively stable in the stated minimal flavour completion unless an ultraviolet boundary condition is tuned or additional matter cancels Eq.~\eqref{eq:vlq-kinetic-mixing}, but the condition above is easily satisfied by the radiative value of Eq.~\eqref{eq:kinetic-mixing-estimate}, so that freeze-out and hard production are controlled by the $B_1+B_2-2B_3$ charges alone. It does not, however, guarantee that the topological channel dominates the $\chi_2$ width: the unmixed theory has no lepton coupling at all, so $\chi_2\to\chi_1\ell^+\ell^-$, induced by $\epsilon_\gamma\propto g_X$, and $\chi_2\to\chi_1\pi^0\gamma$, induced by the topological operator with coefficient $\propto g_X^2/m_X^2$, are both effectively loop-sized and scale in the same way with the gauge couplings. Their ratio depends on the mass splitting alone, and we show in Sec.~\ref{sec:chi2-decays} that the leptonic mode dominates below $\Delta m_\chi\simeq180$~MeV and the topological one above. This coexistence, with a visible topological branching fraction of order unity, is specific to leptophobic completions; a gauged lepton number, as in $B-3L_\tau$, would open a tree-level invisible channel that swamps the topological decay.

The Higgs portal has complementary effects.  The combination $\lambda_{H1}-\lambda_{H2}$ contributes to the common pNGB mass as shown in Eq.~\eqref{eq:higgs-portal-mass}, but it cannot split $\chi_1$ and $\chi_2$ while $U(1)_D$ is exact.  The $\phi_2$ condensate shifts the coefficient of $H^\dagger H$ by $\lambda_{H2}v_D^2/2$ and produces the $h$--$\rho$ mixing entry $\lambda_{H2}v_{\rm EW}v_D$.  If kinematically open, the same portal permits Higgs decays to dark scalars, so Higgs-width and scalar-mixing bounds constrain it.  In the phenomenological analysis below we assume $\epsilon_Y$ and the Higgs-portal couplings are small enough not to affect freeze-out or the hard-production rates; their boundary values are additional model parameters, not predictions of $g_X$, $m_X$, and $Q_\chi$.

\section{Thermal freeze-out above the QCD scale}
\label{sec:freezeout}

The thermal relic abundance probes the portal at center-of-mass energies $\sqrt{s}\simeq m_{\chi_1}+m_{\chi_2}$, rather than at the much lower freeze-out temperature. In the phenomenologically relevant regime of GeV-scale dark matter, this momentum transfer lies at or above the QCD scale, where neither the chiral description in terms of pions nor the corresponding local topological-portal interaction is under perturbative control. Moreover, when the mediator is not parametrically heavier than the dark matter, and especially near the $X_\mu$ resonance, it cannot consistently be integrated out. 

Therefore, in this Section we compute freeze-out directly in the full weakly coupled model, using quark degrees of freedom and retaining the complete $X_\mu$ propagator. Besides providing a controlled description of both resonant and non-resonant annihilation, this treatment expresses the relic-density prediction in terms of the same masses and couplings that determine the collider phenomenology, while the topological portal remains the appropriate low-energy description of processes governed by the much smaller mass splitting. We implement the standard coannihilation formalism, in which the equilibrium
weights of all chemically coupled states enter a single effective annihilation
rate~\cite{Griest:1990kh, Edsjo:1997bg, Baker:2015qna}.

To summarise the results of the previous Section, the scenario we consider is effectively described by two real scalar dark-sector states $\chi_1$ and $\chi_2$ interacting through a massive vector boson $X_\mu$ coupled to the $U(1)_{B_1+B_2-2B_3}$ current. The relevant Lagrangian for everything that follows is
\begin{align}
    \mathcal L &=
    \mathcal L_{\rm SM}
    - \frac14 X_{\mu\nu}X^{\mu\nu} +\frac{1}{2} m_X^2 X_\mu X^\mu
    + \frac12 (\partial_\mu \chi_1)^2
    + \frac12 (\partial_\mu \chi_2)^2 -\frac{1}{2} m_{\chi_1}^2 \chi_1^2 -\frac{1}{2} m_{\chi_2}^2 \chi_2^2\nonumber\\
    &+ g_X X_\mu J_X^\mu
    + g_X Q_\chi
    X_\mu
    \left(
        \chi_1 \partial^\mu \chi_2
        -
        \chi_2 \partial^\mu \chi_1
    \right)\,.
\end{align}
The SM current is
\begin{equation}
    J_X^\mu
    =
    \sum_q Q_q \bar q \gamma^\mu q\,,
\end{equation}
where $q$ runs over the kinematically accessible SM quarks.
For the concrete model of \S \ref{sec:UV-concrete},
\begin{equation}
    Q_u=Q_d=Q_c=Q_s=x_q,
    \qquad
    Q_t=Q_b=-2x_q,
    \qquad
    Q_\chi = x_\phi\,,
\end{equation}
and, in what follows, we define $g_X$ by setting $x_q=1$. Therefore, the model phenomenology studied below is completely fixed in terms of five parameters:
\begin{equation*}
    \{g_X\,,Q_\chi\,,m_X\,,m_{\chi_1}\,,m_{\chi_2}\}\,.
\end{equation*}
The process relevant for freeze-out is $\chi_1\chi_2$ coannihilation into the accessible quark pairs through the $X_\mu$ exchange
    $\chi_1 (p_1) \chi_2 (p_2) \to X^\ast \to q(k_1) \bar q(k_2)$,
with the corresponding particle four-momenta shown in the parentheses.
The tree-level amplitude is given by
\begin{equation}
    \mathcal M
    =
    \frac{
        g_X^2 Q_\chi Q_q
    }{
        s-m_X^2+i m_X \Gamma_X
    }
    (p_1-p_2)_\mu
    \,
    \bar u(k_1)\gamma^\mu v(k_2)\,,
\end{equation}
where
$s=(p_1+p_2)^2$.
After summing over final-state spins and integrating over the scattering angle, one finds the following annihilation cross-section
\begin{equation}
    \sigma_q(s)
    =
    \frac{
        N_c^q g_X^4 Q_q^2 Q_\chi^2
    }{
        12\pi
    }
    \frac{
        (s+2m_q^2)
        \sqrt{\lambda(s,m_{\chi_1}^2,m_{\chi_2}^2)}
        \sqrt{\lambda(s,m_q^2,m_q^2)}
    }{
        s^2
        \left[
            (s-m_X^2)^2
            +
            m_X^2\Gamma_X^2
        \right]
    },
\end{equation}
where
$\lambda(s,a,b) = s^2+a^2+b^2-2sa-2sb-2ab$
is the K\"all\'en function. Finally, the total cross section is obtained by summing over all kinematically accessible quark flavours
\begin{equation}
    \sigma(s)=\sum_q \sigma_q(s)\,.
\end{equation}
To make the main parametric dependences of the relic abundance transparent, it is useful to distinguish two kinematic regimes. Away from the mediator pole, the annihilation rate admits a non-relativistic velocity expansion and its dependence on the masses and couplings can be read off analytically. Near the $X_\mu$ resonance, this expansion breaks down and the thermally averaged Breit--Wigner enhancement must instead be retained. We discuss the non-resonant and resonant regimes in turn below.


\subsection{Non-resonant freeze-out}

In the region away from resonance, defined by the following condition
\begin{equation}
    |m_X^2-(m_{\chi_1}+m_{\chi_2})^2|
    \gg
    m_X\Gamma_X\,,
    \label{eq:condition_nonres}
\end{equation}
the annihilation cross section may be expanded in the non-relativistic limit.
Using
\begin{equation}
    s
    =
    (m_{\chi_1}+m_{\chi_2})^2
    +
    m_{\chi_1}m_{\chi_2}v_{\rm rel}^2
    +
    \mathcal O(v_{\rm rel}^4)\,, \qquad     \lambda(s,m_{\chi_1}^2,m_{\chi_2}^2)
    =
    4m_{\chi_1}^2 m_{\chi_2}^2  v_{\rm rel}^2 +
    \mathcal O(v_{\rm rel}^4)\, ,
\end{equation}
one finds the expected $p$-wave scaling $\sigma v_{\rm rel} = b\,v_{\rm rel}^2$, with the coefficient $b$ reading
\begin{align}
    b
    =
    \sum_q
    \frac{
        N_c^q g_X^4 Q_q^2Q_\chi^2 m_{\chi_1}m_{\chi_2}
    }{
        6\pi \left[
            (m_{\chi_1}+m_{\chi_2})^2-m_X^2
        \right]^2
    }
    \left(
        1+\frac{2m_q^2}{(m_{\chi_1}+m_{\chi_2})^2}
    \right)
    \sqrt{
        1-\frac{4m_q^2}{(m_{\chi_1}+m_{\chi_2})^2}
    }\,,
    \label{eq:b_nonres}
\end{align}
and the thermal average giving $\langle \sigma_{12} v_{\rm rel}\rangle = b \langle v_{\rm rel}^2\rangle$, with
\begin{equation}
    \langle v_{\rm rel}^2\rangle
    =
    \frac{3T}{\mu}\,,
    \qquad
    \mu=\frac{m_{\chi_1}m_{\chi_2}}{m_{\chi_1}+m_{\chi_2}}\,.
\end{equation}
Therefore
\begin{equation}
    \langle \sigma_{12}v\rangle
    =
    \frac{3Tb}{\mu}\,,
\end{equation}
where we used $\sigma_{12}$ to denote the cross section from the $\chi_1-\chi_2$ collisions. In the coannihilation regime, the total dark matter abundance is controlled by the effective cross section
\begin{equation}
    \langle \sigma_{\rm eff}v\rangle
    =
    \sum_{ij}
    r_i r_j
    \langle \sigma_{ij}v\rangle,
\end{equation}
where
\begin{equation}
    r_i
    =
    \frac{
        n_i^{\rm eq}
    }{
        n_1^{\rm eq}+n_2^{\rm eq}
    }.
\end{equation}
Since only the off-diagonal process contributes, $\sigma_{11}=\sigma_{22}=0$, one obtains
\begin{equation}
    \langle \sigma_{\rm eff}v\rangle
    =
    2r_1r_2
    \langle \sigma_{12}v\rangle\,.
\end{equation}
For the mass-splitting parameter defined as $\Delta = m_{\chi_2}/m_{\chi_1}-1>0$,
the equilibrium fractions are
\begin{equation}
    r_1
    =
    \frac{
        1
    }{
        1+(1+\Delta)^{3/2}e^{-x\Delta}
    }\,,\quad
    r_2
    =
    \frac{
        (1+\Delta)^{3/2}e^{-x\Delta}
    }{
        1+(1+\Delta)^{3/2}e^{-x\Delta}
    }\,,
\end{equation}
with $x=m_{\chi_1}/{T}$.

Provided that conversion processes such as $\chi_2 f\leftrightarrow\chi_1 f$ remain faster than the Hubble expansion around freeze-out, they enforce $n_i/n\simeq n_i^{\rm eq}/n^{\rm eq}\equiv r_i$, allowing the coupled two-species system to be reduced to a single Boltzmann equation for the total yield $Y=(n_1+n_2)/s_{\rm ent}$. Assuming entropy conservation and radiation domination, this equation is
\begin{equation}
    \frac{{\rm d}Y}{{\rm d}x}
    =
    -\frac{s_{\rm ent}\langle\sigma_{\rm eff}v\rangle}{Hx}
    \left(Y^2-Y_{\rm eq}^2\right),
\end{equation}
where $s_{\rm ent}$ is the entropy density and $H$ is the Hubble rate. In the non-relativistic limit, the total equilibrium density is proportional to $g_{\rm eff}(x)x^{3/2}e^{-x}$. Applying the standard sudden-freeze-out criterion, namely that the departure of $Y$ from $Y_{\rm eq}$ becomes of order $Y_{\rm eq}$, gives the iterative relation
\begin{equation}
    x_f
    =
    \ln
    \left[
        \frac{
            0.038\,
            g_{\rm eff}(x_f)\,
            M_{\rm Pl}\,
            m_{\chi_1}\,
            \langle \sigma_{\rm eff}v\rangle(x_f)
        }{
            \sqrt{
                g_\ast(x_f)x_f
            }
        }
    \right],
\end{equation}
where $M_{\rm Pl}$ is the Planck mass and we have used $H=1.66\sqrt{g_*(T)}\,T^2/M_{\rm Pl}$. Here, $g_*(T)$ denotes the effective number of relativistic degrees of freedom contributing to the total energy density, $\rho_{\rm rad}=\pi^2 g_*(T)T^4/30$. For two real scalars, each with one internal degree of freedom, the effective number of equilibrium degrees of freedom is
\begin{equation}
    g_{\rm eff}(x)
    =
    1+(1+\Delta)^{3/2}e^{-x\Delta}.
\end{equation}
The first term is the contribution of $\chi_1$, while the second contains both the non-relativistic phase-space factor $(1+\Delta)^{3/2}$ and the Boltzmann suppression of the heavier state.

After freeze-out, $Y_{\rm eq}$ rapidly becomes negligible. The entropy density is
\begin{equation*}
    s_{\rm ent}=\frac{2\pi^2}{45}g_{*s}(T)T^3,
\end{equation*}
where $g_{*s}(T)$ is the effective number of relativistic degrees of freedom contributing to the entropy density. Integrating the Boltzmann equation from $x_f$ to the present and neglecting the subleading boundary term $Y^{-1}(x_f)$ gives
\begin{equation}
    Y_\infty^{-1}
    =
    \sqrt{
        \frac{\pi}{45}
    }
    M_{\rm Pl}m_{\chi_1}
    \int_{x_f}^\infty
    dx\,
    \frac{
        \sqrt{g_\ast}\,
        \langle \sigma_{\rm eff}v\rangle
    }{
        x^2
    }\,,
\end{equation}
where we have neglected the temperature variation of the relativistic degrees of freedom during freeze-out and used $g_{*s}(T)\simeq g_*(T)$. Finally, after the heavier state has decayed into $\chi_1$, the present mass density is $\rho_\chi=m_{\chi_1}s_0Y_\infty$. Dividing by the critical density therefore gives
\begin{equation}
    \Omega_\chi h^2
    =
    2.75\times10^8
    \left(
        \frac{m_{\chi_1}}{\rm GeV}
    \right)
    Y_\infty\,.
\end{equation}
The mass degenerate limit
%
    $m_{\chi_1}=m_{\chi_2}=m_\chi$
%
implies
%
    $r_1=r_2=1/2$,
%
and consequently
\begin{equation}
    \langle \sigma_{\rm eff}v\rangle
    =
    \frac12
    \langle \sigma_{12}v\rangle.
\end{equation}
\noindent For a fixed mediator mass, in general, there are two solutions for $m_\chi$ that reproduce the correct relic abundance, lying on either side of the \textit{pole} defined by the $m_X=2m_\chi$ line. This is due to the quadratic dependence of the denominator in Eq.~\eqref{eq:b_nonres}, which ultimately determines the value of $\Omega_\chi h^2$. Interestingly, as the masses increase both solutions approach this pole and the condition for the non-resonant freeze-out in Eq.~\eqref{eq:condition_nonres} is necessarily violated. Thus, the non-resonant velocity expansion ceases to be reliable and the full resonant treatment becomes essential. We study this regime next.    


\subsection{Resonant freeze-Out}

In the region near resonance,
\begin{equation}
    m_X^2
    \simeq
    (m_{\chi_1}+m_{\chi_2})^2,
\end{equation}
the velocity expansion used above breaks down because the annihilation rate varies rapidly across the thermally populated range of centre-of-mass energies. It is therefore necessary to retain the full thermal average. Writing $M_\chi=m_{\chi_1}+m_{\chi_2}$, we introduce
\begin{align}
    s
    &=
    M_\chi^2(1+z)\,,\quad\quad
    \delta
    =
    \frac{
        m_X^2-M_\chi^2
    }{
        M_\chi^2
    }\,,\quad\quad
    \gamma
    =
    \frac{
        m_X\Gamma_X
    }{
        M_\chi^2
    }\,,
\end{align}
where $z$ measures the kinetic energy above threshold, $\delta$ the position of the pole relative to threshold, and $\gamma$ its dimensionless width. The exact thermal average is
\begin{equation}
    \langle\sigma_{12} v_{\rm rel}\rangle = \frac{\sum_q N_c^q g_X^4 Q_q^2 Q_\chi^2}{96 \pi m_{\chi_1}^2 m_{\chi_2}^2 T\,{\rm K}_2(m_{\chi_1}/T) {\rm K}_2(m_{\chi_2}/T)} \int_{s_{\rm th}}^{\infty} {\rm d}s \frac{(s+2m_q^2)\lambda_{i}^{3/2} \lambda_q^{1/2} {\rm K}_1(\sqrt{s}/T)}{s^{5/2} [(s-m_X^2)^2+m_X^2 \Gamma_X^2]}\,,
\end{equation}
with $s_{\rm th}=M_\chi^2$, $\lambda_i = \lambda(s,m_{\chi_1}^2,m_{\chi_2}^2)$, and $\lambda_q = \lambda(s,m_q^2,m_q^2)$. We use this expression for the numerical calculation. To expose the parametric behaviour near a narrow resonance, we take $M_\chi/T\gg1$ and $z,\delta\ll1$, for which
\begin{align}
    {\rm K}_{1}\left(\frac{\sqrt{s}}{T}\right) &= {\rm K}_{1}\left(\frac{M_\chi}{T} \sqrt{1+z}\right)\simeq \sqrt{\frac{\pi\, T}{2 (M_\chi) \sqrt{1+z}}} e^{-\frac{M_\chi}{T}\sqrt{1+z}}\,,\\
    {\rm K}_2\left(\frac{m}{T}\right) &\simeq \sqrt{\frac{\pi\,T}{2m}} e^{-\frac{m}{T}}\,,
\end{align}
Keeping the leading terms in this expansion gives
\begin{align}
    \langle\sigma_{12} v_{\rm rel}\rangle &\simeq \frac{\sum_q \sqrt{2} N_c^q g_X^4 Q_q^2 Q_\chi^2}{12 \pi^{3/2}M_\chi^5} \left(\frac{M_\chi}{T}\right)^{3/2}(M_\chi^2+2m_q^2)\nonumber\\
    &\times(M_\chi^2-4m_q^2)^{1/2} \int_0^\infty {\rm d}z \frac{z^{3/2} e^{-M_\chi z/(2T)}}{\gamma^2+(z-\delta)^2}\,.
\end{align}
The pole lies inside the physical integration region only for $\delta>0$. If it is also narrow compared with the scales over which the remaining integrand varies, the Breit--Wigner factor may be replaced by
\begin{equation}
    \frac{1}{(z-\delta)^2+\gamma^2}
    \to
    \frac{\pi}{\gamma}\delta_{\rm D}(z-\delta)\,,
\end{equation}
where $\delta_{\rm D}$ denotes the Dirac delta distribution. The integral then becomes
\begin{equation}
    \int_0^\infty {\rm d}z \frac{z^{3/2} e^{-M_\chi z/(2T)}}{\gamma^2+(z-\delta)^2}
    \simeq \frac{\pi}{\gamma} \delta^{3/2} e^{-M_\chi\delta/(2T)}\,.
\end{equation}

\noindent Thus, up to the more slowly varying prefactors and the subsequent integration of the Boltzmann equation, the resonant annihilation rate scales as $\delta^{3/2}e^{-M_\chi\delta/(2T)}/\gamma$. The corresponding inverse scaling of the relic abundance, evaluated around freeze-out, is
\begin{equation}
    \Omega_\chi h^2
    \propto
    \frac{
        \gamma
    }{
        \delta^{3/2}
    }
    \exp\!\left(\frac{M_\chi\delta}{2T_f}\right)\,.
\end{equation}
For nearly degenerate states, $M_\chi\simeq2m_\chi$ and $x_f=m_\chi/T_f$, the exponential reduces to $e^{x_f\delta}$.

The width of the $X_\mu$ gauge boson plays an important role through the parameter $\gamma$. It is given by
\begin{equation}
    \Gamma_X
    =
    \sum_q \Gamma(X\to q\bar q)
    +
    \Gamma(X\to \chi_1\chi_2),
\end{equation}
with
\begin{equation}
    \Gamma(X\to q\bar q)
    =
    \frac{
        N_c^q g_X^2 Q_q^2 m_X
    }{
        12\pi
    }
    \left(
        1+\frac{2m_q^2}{m_X^2}
    \right)
    \sqrt{
        1-\frac{4m_q^2}{m_X^2}
    },
\end{equation}
valid for $m_X>2m_q$, and
\begin{equation}
    \Gamma(X\to \chi_1\chi_2)
    =
    \frac{
        g_X^2 Q_\chi^2
    }{
        48\pi m_X^5
    }
    \lambda^{3/2}(m_X^2,m_{\chi_1}^2,m_{\chi_2}^2),
\end{equation}
valid for $m_X>m_{\chi_1}+m_{\chi_2}$.
In the mass degenerate limit $m_{\chi_1}=m_{\chi_2}=m_\chi$, the dark-sector contribution to the width simplifies to
\begin{equation}
    \Gamma(X\to \chi\chi)
    =
    \frac{
        g_X^2 Q_\chi^2 m_X
    }{
        48\pi
    }
    \left(
        1-\frac{4m_\chi^2}{m_X^2}
    \right)^{3/2}.
\end{equation}

\subsection{Revisiting direct detection}
\label{sec:DD}

Since $X_\mu$ couples to the dark sector only through the off-diagonal current $\propto \chi_1 \partial_\mu \chi_2 - \chi_2 \partial_\mu \chi_1$, there is no $\chi_1\chi_1X$ vertex and a nucleus can be struck only through the transition $\chi_1N\to\chi_2N$. Equipped with our concrete UV model, we here study the expectations for direct detection signals due to up-scattering through this inelastic channel, which depends significantly on the dark pion mass splitting.

The vertex $g_XQ_\chi(p_1+p_2)^\mu\simeq2m_{\chi_1}g_XQ_\chi\,\delta^{\mu0}$ is not velocity suppressed, since a derivative acting on a non-relativistic field produces the mass rather than the momentum; what closes the channel is kinematics. Energy conservation in the centre-of-mass frame, $\tfrac12\mu_Nv^2=\tfrac12\mu_Nv'^2+\Delta m_\chi$, with $\mu_N$ the DM--nucleus reduced mass and $v\le v_{\max}=v_{\rm esc}+v_\odot\simeq800~{\rm km/s}$~\cite{Baxter:2021pqo}, requires
\begin{equation}
    \Delta m_\chi\le\tfrac12\,\mu_N v_{\max}^2\simeq3.5~{\rm keV}\times\frac{\mu_N}{\rm GeV}\,,
    \qquad\text{equivalently}\qquad
    v\ge v_{\rm th}=\sqrt{2\Delta m_\chi/\mu_N}\,.
    \label{eq:DD-kinematics}
\end{equation}
For GeV dark pions $\mu_N\simeq m_{\chi_1}$ and the bound is a few keV; for $m_{\chi_1}\gg m_N$ it saturates at $\tfrac12m_Nv_{\max}^2\simeq0.4$~MeV for xenon and remains below $1$~MeV for any nucleus. 

The mass splittings of interest here, $\Delta m_\chi\gtrsim m_{\pi^0}$, exceed Eq.~\eqref{eq:DD-kinematics} by four to five orders of magnitude for every DM mass and target: the threshold speed is $v_{\rm th}\simeq0.5c$ for $m_{\chi_1}=1$~GeV and still $0.05c$ for $m_{\chi_1}\to\infty$, and even the solar potential well, which accelerates DM to $\sim1400$~km/s, supplies only $\sim10$~keV per GeV of $\mu_N$. Direct detection through the portal is therefore essentially forbidden, not just suppressed~\cite{Tucker-Smith:2001myb}. The exothermic channel $\chi_2N\to\chi_1N$ is absent as well, because the heavier $\chi_2$ particle decays long before today (Sec.~\ref{sec:chi2-decays}). What remains, for this regime of masses and splittings,
is elastic scattering through the Higgs portal,
\begin{equation}
    \sigma^{\rm el}_n\simeq\frac{\lambda_{H1}^2f_N^2\,\mu_n^2m_N^2}{4\pi m_h^4m_{\chi_1}^2}
    \simeq6\times10^{-44}~{\rm cm^2}\left(\frac{\lambda_{H1}}{10^{-2}}\right)^{2}\left(\frac{3~{\rm GeV}}{m_{\chi_1}}\right)^{2},
    \label{eq:DD-higgs}
\end{equation}
with $f_N\simeq0.3$ and $\mu_n$ the DM--nucleon reduced mass, which current low-mass searches~\cite{LZ:2024zvo,XENON:2025vwd,PandaX:2024qfu} restrict to $\lambda_{H1}\lesssim10^{-2}$ for $m_{\chi_1}\simeq3$--$5$~GeV, and one-loop double-$X$ exchange, which generates only $U(1)_D$-symmetric operators suppressed by $(g_\chi g_q/4\pi)^2$ and by quark masses, far below any foreseeable sensitivity.

When the transition is open, Eq.~\eqref{eq:DD-kinematics} implies a recoil spectrum concentrated at $E_R\simeq(\mu_N/m_N)\,\Delta m_\chi$ with no events at low energy~\cite{Tucker-Smith:2001myb,Bramante:2016rdh}. LZ has recently extended its nuclear-recoil window to $270$~keV to search for this signature and reports a single event at $248$~keV~\cite{LZ:2026ext}, interpreted as inelastic scattering of TeV-scale DM with $\Delta m_\chi\simeq300$~keV~\cite{Su:2026rwz,Fan:2026kxx,Freese:2026sga,Wu:2026nhi} (see also~\cite{Pospelov:2026ewn}). By Eq.~\eqref{eq:DD-kinematics} such searches are sensitive only to $\Delta m_\chi\lesssim0.4$~MeV and $m_{\chi_1}\gtrsim0.4$~TeV; in that regime the very same $U(1)_X$ portal, with a thermal cross section, accounts for the LZ event~\cite{Davighi:2026lz}. For the GeV dark pions of this paper, direct detection is blind, and the portal has to be probed at colliders.

\subsection{Revisiting indirect detection}
\label{sec:ID}

The same structure suppresses indirect detection. The process that sets the relic abundance, $\chi_1\chi_2\to X^*\to q\bar q$, is $p$-wave and requires a $\chi_2$, none of which survives today; the dark matter is a single real scalar without a tree-level coupling of its own to $X$. Tree-level annihilation of $\chi_1$ pairs needs two insertions, $\chi_1\chi_1\to X^*X^*\to q\bar q\,q'\bar q'$. For $m_X\gg m_{\chi_1}$ and massless quarks the $s$-wave rate is, up to an $\mathcal O(1)$ factor from the $\chi_2$-exchange diagrams,
\begin{equation}
    \langle\sigma v\rangle_{\chi_1\chi_1}
    \simeq\frac{\big(\sum_qN_c^qx_q^2\big)^2\,g_\chi^4g_q^4\,m_{\chi_1}^6}{864\,\pi^5\,m_X^8}
    =\frac{\sum_qN_c^qx_q^2}{144\,\pi^4}\,(g_\chi g_q)^2\Big(\frac{m_{\chi_1}}{m_X}\Big)^{\!4}\Big(1-\frac{4m_{\chi_1}^2}{m_X^2}\Big)^{\!2}\,b\,,
    \label{eq:ID-4q}
\end{equation}
with $b$ the coannihilation coefficient of Eq.~\eqref{eq:b_nonres}: two extra portal couplings, a four-body phase space and $(m_{\chi_1}/m_X)^4$ suppress it by $\sim10^{-10}$ relative to the thermal rate for $g_\chi g_q\sim10^{-2}$ and $m_X\sim4m_{\chi_1}$. 

Closing the two $X$ lines on a quark line gives instead the one-loop $2\to2$ process $\chi_1\chi_1\to q\bar q$. Since the initial state has $J=0$ and $X$ couples to vector currents, its amplitude is helicity suppressed, $\propto m_q$, as for any scalar annihilating into fermions, and
\begin{equation}
    \langle\sigma v\rangle_{\chi_1\chi_1\to q\bar q}
    \simeq\frac{N_c^q\,(g_\chi g_q)^4\,m_q^2\,\beta_q^3}{1024\,\pi^5\,m_X^4}
    \label{eq:ID-loop}
\end{equation}
up to an $\mathcal O(1)$ loop function, with $\beta_q$ the quark velocity. It exceeds Eq.~\eqref{eq:ID-4q} by $\sim10^{-2}(m_q/m_{\chi_1})^2(m_X/m_{\chi_1})^4$, i.e.\ for $m_X\gtrsim4m_{\chi_1}$, while the helicity-unsuppressed $gg$ and $\gamma\gamma$ final states arise only at two loops. 

Along the relic-density contours of Fig.~\ref{fig:parameter-scans} all these channels stay below $10^{-30}~{\rm cm^3\,s^{-1}}$ for $m_X\gtrsim1.2\,m_{\chi_1}$, including the three-body $\chi_1\chi_1\to XX^*$ that opens for $m_X<2m_{\chi_1}$. The Higgs portal adds $\langle\sigma v\rangle_h=\sum_fN_c^f\lambda_{H1}^2m_f^2\beta_f^3/(4\pi m_h^4)\simeq6\times10^{-30}~{\rm cm^3\,s^{-1}}\,(\lambda_{H1}/10^{-2})^2$ at $m_{\chi_1}=5$~GeV, with $\beta_f$ the fermion velocity. All are more than three orders of magnitude below the Fermi-LAT dwarf-spheroidal~\cite{Fermi-LAT:2015att} and Planck~\cite{Slatyer:2015jla,Planck:2018vyg} sensitivities, $\langle\sigma v\rangle\sim10^{-27}$--$10^{-26}~{\rm cm^3\,s^{-1}}$ for GeV-scale DM annihilating into hadrons. The exception is $m_X<m_{\chi_1}$, where $\chi_1\chi_1\to XX$ opens with an $s$-wave cross section of thermal size, $\pi\alpha_\chi^2/m_{\chi_1}^2$, and the CMB excludes it~\cite{Slatyer:2015jla}; we do not consider that region.

The one environment in which the transition of Eq.~\eqref{eq:DD-kinematics} is open is a neutron star, where infalling DM reaches $v\simeq0.6c$. Since $\sigma_n$ exceeds the geometric threshold of $\sim10^{-45}~{\rm cm^2}$ by many orders of magnitude, every incident $\chi_1$ up-scatters and is captured, and the deposited kinetic energy, together with the $\Delta m_\chi$ released in the subsequent $\chi_2\to\chi_1\pi^0\gamma$ decay, heats old neutron stars to the $\sim2000$~K level~\cite{Baryakhtar:2017dbj,Bell:2018pkk}, which is an observation potentially within the reach of infrared telescopes.\footnote{Because $\chi_1$ does not annihilate, the captured population accumulates. Whether non-annihilating bosonic DM in neutron stars collapses to a black hole~\cite{Kouvaris:2011fi,McDermott:2011jp} is controlled by its self-interactions, here of order $m_{\chi_1}^2/f_D^2$; we leave this question to future work.}

\section{Collider phenomenology} \label{sec:colliders}

Having re-evaluated the relic abundance calculations using the full microscopic completion of the topological portal, we now have a target region in parameter space for which to study phenomenology. 
The model features a $Z^\prime$ gauge boson, the vector $X_\mu$ of Sec.~\ref{sec:topological}, coupled to the light-family baryon current, as required to match onto the topological portal. In \S \ref{sec:UV-concrete} we set out a concrete model that does this, wherein the $Z^\prime$ couples to the $B_1+B_2-2B_3$ current. In this Section we explore the rich collider phenomenology of this $Z^\prime$. Motivated by fitting the relic abundance, we are interested in regions where the mass of the dark matter is several GeV, for which the $Z^\prime$ mediator mass is tens of GeV. 

Since the searches we consider below produce the dark pair $\chi_1\chi_2$, we first establish in Sec.~\ref{sec:chi2-decays} how, and how fast, the excited state $\chi_2$ decays: this decides whether the dark states appear as missing energy or as displaced decays, and it is where the infrared topological operator, rather than the ultraviolet completion, controls the phenomenology.

\subsection{Decays of the excited dark state}
\label{sec:chi2-decays}

The exact parity $P_D$ renders the lighter $\chi_1$ particle stable, while the heavier $\chi_2$ decays through the off-diagonal dark current with a momentum release set by the splitting $\Delta m_\chi\equiv m_{\chi_2}-m_{\chi_1}\ll m_{\chi_1}$ of Eq.~\eqref{eq:dark-masses}. This is far below the scale of production and freeze-out, and the appropriate description is the low-energy one of Sec.~\ref{sec:uv-general}: the quark current in Eq.~\eqref{eq:Ldim6} matches onto the topological operator~\eqref{eq:2to2}, whose coefficient in the one-condensate model, Eq.~\eqref{eq:single-condensate-mass} with $Q_\chi=3$, reads
\begin{equation}
    C_{\pi\gamma}
    =\frac{N e}{16\pi^2 f_\pi f_D^2}
    =\frac{N e\,Q_\chi^2}{4\pi^2 f_\pi}\,\frac{g_X^2}{m_X^2}\,,
    \qquad \text{where~~} N=n_c x_q=3\,,
    \label{eq:chi2-topological}
\end{equation}
so that $f_D=m_X/(6g_X)\simeq170~{\rm GeV}\,(m_X/10~{\rm GeV})(10^{-2}/g_X)$ lies well above the dark-matter mass.

The topological portal operator induces the decay $\chi_2\to\chi_1\pi^0\gamma$ for $\Delta m_\chi>m_{\pi^0}$, with partial width
\begin{equation}
    \Gamma_{\pi^0\gamma}
    =\frac{C_{\pi\gamma}^2}{1536\pi^3m_{\chi_2}^3}
    \int_{m_{\pi^0}^2}^{\Delta m_\chi^2}\!ds\,
    \frac{(s-m_{\pi^0}^2)^3}{s^2}\,
    \lambda^{3/2}\!\left(m_{\chi_2}^2,m_{\chi_1}^2,s\right)
    \;\simeq\;
    \frac{C_{\pi\gamma}^2}{1680\pi^3}\,\Delta m_\chi^7\,,
    \label{eq:chi2-pigamma-width}
\end{equation}
where $\lambda$ is the K\"all\'en function of Sec.~\ref{sec:freezeout} and the last approximation holds if we assume a mass hierarchy $m_{\pi^0}\ll\Delta m_\chi\ll m_{\chi_1}$. 
We see this depends very steeply on the mass splitting, as $\Gamma_{\pi^0 \gamma} \propto \Delta m_\chi^7$. 
This comes from the two spacetime derivatives that appear in the topological vertex, as well as the three-body phase space that we integrate over. Close to threshold, evaluating the exact expression is numerically much smaller than its asymptotic form, by a factor of six at $\Delta m_\chi=300$~MeV (and so we should not use the approximation in this regime).

The topological portal operator contains couplings not just to $\pi^0 \gamma$, but to other QCD final states, with relative couplings fixed by $SU(3)$-flavour group theory factors.
In particular, the same current opens $\chi_2\to\chi_1\pi^+\pi^-\pi^0$ above $414$~MeV and $\chi_2\to\chi_1\eta\gamma$ above $m_\eta$. Both are strongly phase-space suppressed for $\Delta m_\chi\lesssim500$~MeV, the range we consider, and are neglected below. 
The result is that we slightly overestimate the $\chi_2$ lifetime at the upper end of this range. Finally, if we were increase the mass-splitting further, towards the $\omega(782)$ region, the leading-order chiral description would no longer be reliable.

The second important channel to consider is leptonic. For the concrete $B_1+B_2-2B_3$ completion we set out in \S \ref{sec:UV-concrete}, kinetic mixing between the $Z^\prime$ and the photon is radiatively generated at 1-loop by the vector-like quarks (needed to complete the CKM matrix), as per Eq.~\eqref{eq:kinetic-mixing-estimate}. At low energies this induces couplings $\cL \supset \epsilon_\gamma eX_\mu J^\mu_{\rm em}$, with $\epsilon_\gamma\simeq c_W\epsilon_Y$. Integrating out $X$, we obtain
\begin{equation}
    \mathcal L_{\ell\ell}
    =G_\ell\left(\chi_1\partial_\mu\chi_2-\chi_2\partial_\mu\chi_1\right)\bar\ell\gamma^\mu\ell\,,
    \qquad
    G_\ell=\frac{g_XQ_\chi\,\epsilon_\gamma e}{m_X^2}\,,
    \qquad
    \Gamma_{\ell\ell}\simeq\frac{G_\ell^2}{60\pi^3}\,\Delta m_\chi^5
    \label{eq:chi2-dilepton}
\end{equation}
for $m_\ell\ll\Delta m_\chi\ll m_{\chi_1}$; our numerical results use the full lepton-mass dependence.\footnote{Explicitly, $d\Gamma_{\ell\ell}/ds=G_\ell^2\,\lambda^{3/2}(m_{\chi_2}^2,m_{\chi_1}^2,s)\,(1-4m_\ell^2/s)^{1/2}(1+2m_\ell^2/s)/(192\pi^3m_{\chi_2}^3)$ for $4m_\ell^2<s<\Delta m_\chi^2$.} Neutrinos couple to $X$ only through $Z$--$X$ mixing, suppressed by $m_X^2/m_Z^2$, and $\chi_2\to\chi_1\nu\bar\nu$ is negligible. This is where leptophobia matters: in a gauged $B-3L_\tau$ completion, for instance, $x_{\nu_\tau}=-9$ in our normalization and the tree-level invisible width, $\Gamma_{\nu\bar\nu}\simeq G_\nu^2\Delta m_\chi^5/(120\pi^3)$ with $G_\nu=g_X^2Q_\chi x_{\nu_\tau}/m_X^2$, exceeds the topological one by $\Gamma_{\nu\bar\nu}/\Gamma_{\pi^0\gamma}\simeq14\,[4\pi^2f_\pi/(e\,\Delta m_\chi)]^2\simeq8\times10^3\,(500~{\rm MeV}/\Delta m_\chi)^2$ for $N=Q_\chi=3$. 

Sizeable topological branching fractions are thus a distinctive feature of leptophobic UV realisations of the topological portal, such as through gauging $B_1+B_2-2B_3$; we need couplings to QCD to dominate over couplings to leptons to see these portal-induced decays.

In the completion with a single up-type vector-like quark and vanishing kinetic mixing at some UV scale $\Lambda>M_U$, Eq.~\eqref{eq:kinetic-mixing-estimate} gives $\epsilon_\gamma/g_X\simeq(e/6\pi^2)\ln(\Lambda/M_U)\simeq5\times10^{-3}\ln(\Lambda/M_U)$, a representative radiative value rather than a prediction, as it depends on the ultraviolet boundary condition and on threshold effects. Since $\epsilon_\gamma\propto g_X$, both $C_{\pi\gamma}$ and $G_\ell$ scale as $g_X^2/m_X^2$: the branching ratios depend on $\Delta m_\chi$ alone, while the lifetime carries the entire dependence on the gauge sector, $c\tau_{\chi_2}\propto g_X^{-4}m_X^4$, so that reducing $g_X$ by one order of magnitude lengthens the decay by four. For $\ln(\Lambda/M_U)=1$, including the $e^+e^-$, $\mu^+\mu^-$ and $\pi^0\gamma$ modes with full phase space, we find the branching ratios and decay lengths of Table~\ref{tab:chi2-decays}. Below the pion threshold and up to $\Delta m_\chi\simeq180$~MeV the decay is leptonic; above $200$~MeV the topological mode takes over and the lifetime drops steeply. The benchmark $\Delta=0.05$ and $m_{\chi_1}=10$~GeV of Fig.~\ref{fig:parameter-scans} corresponds to the upper end of the range considered here, where ${\rm Br}(\chi_2\to\chi_1\pi^0\gamma)\simeq1$.
Thus, for this benchmark point and others nearby, the topological portal does indeed dominate the decay of the co-annihilating dark matter particle $\chi_2$.

\begin{table}[t]
  \centering
  \small
  \renewcommand{\arraystretch}{1.15}
  \begin{tabular}{lcccccc}
    \toprule
    $\Delta m_\chi$ [MeV] & 100 & 150 & 180 & 200 & 300 & 500 \\
    \midrule
    ${\rm Br}(\chi_2\to\chi_1\pi^0\gamma)$ & $0$ & $0.006$ & $0.51$ & $0.83$ & $0.99$ & $1.0$ \\
    $c\tau_{\chi_2}\,(10^{-2}/g_X)^{-4}(m_X/10~{\rm GeV})^{-4}$ [m] \quad  & $2\times10^{9}$ & $2\times10^{8}$ & $5\times10^{7}$ & $10^{7}$ & $6\times10^{4}$ & $6\times10^{2}$ \\
    \bottomrule
  \end{tabular}
  \caption{Branching ratio of the topological mode and proper decay length of $\chi_2$ in the one-condensate model ($Q_\chi=3$) with radiative kinetic mixing, $\epsilon_\gamma/g_X=e/6\pi^2$, for $g_X=10^{-2}$ and $m_X=10$~GeV; the remainder of the width is $\chi_2\to\chi_1\ell^+\ell^-$. The decay length scales as $g_X^{-4}m_X^4$, and the last column neglects the $\chi_2 \to \chi_1 3\pi$ decay mode.}
  \label{tab:chi2-decays}
\end{table}

Two conclusions follow for the thermal targets of Fig.~\ref{fig:parameter-scans}, which lie at $m_X\simeq2m_{\chi_1}$ and $10^{-4}\lesssim g_X\lesssim$ a few $\times10^{-2}$. On cosmological time scales $\chi_2$ is short-lived: $\tau_{\chi_2}\lesssim10^4~{\rm s}\,(10^{-3}/g_X)^4(m_X/10~{\rm GeV})^4$ for $\Delta m_\chi\gtrsim150$~MeV, falling as $\Delta m_\chi^{-5}$ to $\Delta m_\chi^{-7}$, so the relic $\chi_2$ population converts into $\chi_1$ before late electromagnetic energy injection is constrained by nucleosynthesis~\cite{Kawasaki:2017bqm}.  On detector scales, in contrast, $\chi_2$ is stable: for $g_X\lesssim10^{-2}$ and $\Delta m_\chi\lesssim500$~MeV its decay length exceeds $600~{\rm m}\,(m_X/10~{\rm GeV})^4$. In the searches of Secs.~\ref{sec:LHC} and~\ref{sec:lepton-colliders} we therefore treat $\chi_1\chi_2$ as missing energy. Displaced or semi-visible $\chi_2\to\chi_1\pi^0\gamma$ decays inside a detector require the upper edge of both ranges, $g_X\gtrsim10^{-2}$ and $\Delta m_\chi\gtrsim0.5$~GeV, where $c\tau_{\chi_2}$ falls to metres; there the long-lived-particle programmes of the LHC and of Belle~II~\cite{Duerr:2019dmv,Alimena:2019zri} complement the missing-energy searches, and we return to this in Sec.~\ref{sec:upsilon}.

\subsection{Unavoidable collider signatures of the topological portal} 
\label{sec:LHC}

Due to the necessary coupling to a light baryon number current, any model that matches onto the topological portal has unavoidable signatures in channels with jets and/or missing energy (thanks to its coupling to dark matter) at the LHC, and in $Z$ decays; the $\Upsilon$ decays discussed below probe in addition the $b$-quark charge, which is fixed by the anomaly-free embedding rather than by the matching onto the portal.  In this Section we explore these signatures and demonstrate their compatibility with fitting the dark matter relic abundance. The experimental measurements turn out to be highly relevant for probing the interesting region, and are set to improve moving forward to the HL-LHC phase.

\subsubsection{Boosted dijet plus photon}

The first search we consider is boosted dijet production, proceeding from $q\bar{q} \to X \to q \bar{q}$. Due to the trigger thresholds, the resonant dijet searches on their own rapidly lose sensitivity at smaller resonance masses. Low-mass dijet searches overcome standard trigger limitations through
trigger-level reconstruction or by requiring hard initial-state radiation,
thereby extending resonance sensitivity well below the conventional dijet
range~\cite{CMS:2016gsl,ATLAS:2018qto,ATLAS:2024qqm,Dobrescu:2013cmh}. Indeed, the CMS search in~\cite{CMS:2019xai} employed such a photon trigger strategy to provide the first hadron collider constraints in the $m_X<50$ GeV region. The collaboration looked for a localized excess in the jet invariant mass spectrum, and in the absence of such a feature, set 95\% C.L. limits on $g_q^\prime$, the coupling strength of resonances decaying to quark pairs. Since our scenario corresponds to the one in the CMS analysis, we recast the bounds $g_X$ by taking into account the appropriate correction for $\mathrm{Br}(X\to q\bar q)$, since the invisible decays are also open. Therefore, the 95\% C.L. limits on $g_X$ is obtained through
\begin{equation}
    g_X = \frac{g_q^\prime}{\sqrt{\mathrm{Br}(X\to q\bar q)}}\,.
\end{equation}
We show the limits of this search by a gray solid line in Fig.~\ref{fig:parameter-scans}, while the HL-LHC projections are shown by a gray dashed line. Other approaches, such as energy correlators~\cite{Ricci:2026fkj}, will become important in the future.

\subsubsection{Monojet}

Another unavoidable collider manifestation of the topological portal UV completion originates from the invisible decay $X\to\chi_1\chi_2$, accompanied by initial-state QCD radiation, resulting in a monojet signature. In order to understand the impact on the model parameter space, we performed a recast of the ATLAS search~\cite{ATLAS:2021kxv} that looked for an energetic jet and missing transverse momentum using $139~{\rm fb}^{-1}$ of $13~{\rm TeV}$ data. Signal events for $pp\to Xj$, followed by $X\to\chi_1\chi_2$, were generated with \textsc{MadGraph5\_aMC@NLO}~\cite{Alwall:2014hca}, requiring $p_T^j>150~{\rm GeV}$, $|\eta_j|<2.4$, and $E_T^{\rm miss}>200~{\rm GeV}$. We have also used $Q_\chi=3$, with $m_{\chi_1} = 10$ GeV (in Fig.~\ref{fig:parameter-scans}~(a,c), and $m_{\chi_1} = 7$ GeV in Fig.~\ref{fig:parameter-scans}~(b)) and $\Delta=0.05$. Away from these values, the experimental constraints inherit a mild dependence on the dark-sector parameters through the branching fractions of $X$. Increasing $Q_\chi$ enhances the invisible width $X\to\chi_1\chi_2$, strengthening the monojet constraint while reducing $\mathrm{BR}(X\to jj)$ and hence weakening the dijet-plus-photon bound. Conversely, increasing $m_{\chi_1}$ or $\Delta$ reduces the available phase space for the invisible decay, or closes it altogether, thereby weakening the monojet sensitivity and strengthening the visible dijet constraint. 

The analysis involved binning of the surviving events in the eight exclusive $E_T^{\rm miss}$ intervals between $200$ GeV and $800~{\rm GeV}$, normalizing to the experimental luminosity and comparing with the background yields. We used the uncertainties reported by ATLAS through an approximate binned Gaussian likelihood, neglecting correlations between bins, and finally constructed a test statistic to determine the $95\%$ C.L. exclusion contour in the $(m_X,g_X)$ plane, shown by a cyan solid line in Fig.~\ref{fig:parameter-scans}. The simulations were performed at the reference coupling $g_X=0.1$, for which the mediator is narrow and the decay-chain implementation of the narrow-width approximation is reliable, and were subsequently rescaled with the coupling. As can be seen from Fig.~\ref{fig:parameter-scans}, the inferred monojet limits consistently reach perturbative couplings for which the extrapolated width satisfies $\Gamma_{X}/m_{X}\lesssim0.3$, shown by a black dash-dotted line. Finally, the projections for the HL-LHC limits are shown by a cyan dashed line.

The green curves in Fig.~\ref{fig:parameter-scans} denote the parameter points that reproduce the observed relic abundance for representative variations of the mass splitting, the dark-matter mass and the dark charge. They are meaningful only for $m_X>m_{\chi_1}$: below it the $s$-wave channel $\chi_1\chi_1\to XX$ opens, dominates freeze-out and is excluded by the CMB (Sec.~\ref{sec:ID}), so this part of the curves is to be disregarded. Each curve then has two non-resonant branches, at $m_X<m_{\chi_1}+m_{\chi_2}$ and at $m_X>m_{\chi_1}+m_{\chi_2}$, joined by the resonant dip. The upper branch, which requires $g_X\gtrsim0.05$, is excluded by the monojet, LEP and $t\to cX$ constraints. The lower branch is not reached by the missing-energy searches, which require an open $X\to\chi_1\chi_2$ channel: for the benchmark $m_{\chi_1}=10$~GeV of panels (a) and (c) it lies at $g_X\simeq0.03$--$0.05$ for $m_X<20$~GeV. Model-independently it is excluded there only by rare $B$ decays, $m_X<4.8$~GeV, and by $\Upsilon(1S)$ decays, $6\lesssim m_X\lesssim10.5$~GeV (Sec.~\ref{sec:upsilon}); the $D$-mixing bound of Sec.~\ref{sec:additional}, which depends on the flavour alignment, covers it up to $m_X\simeq19$~GeV, where the curve bends into the dip. This branch is within reach of the visible FCC-ee search of Sec.~\ref{sec:FCCee-vis} for $12\lesssim m_X\lesssim20$~GeV. The remaining viable region receives an important contribution from resonant annihilation near $m_X\simeq m_{\chi_1}+m_{\chi_2}$. As is apparent from the finite extent of the dips in the relic-density curves, however, this does not require the mediator mass to be precisely tuned to the coannihilation threshold. The resonant region therefore remains a natural part of parameter space and will be tested more extensively by the HL-LHC projections.

\subsubsection{Bottomonium decays}
\label{sec:upsilon}

Below the reach of the LHC searches, the strongest constraint on a vector coupled to the $b$ quark comes from the decays of the $\Upsilon(1S)$, the $b\bar b$ vector ground state with $m_\Upsilon=9.46$~GeV~\cite{Carone:1994aa,Aranda:1998fr,Dobrescu:2014fca,Dobrescu:2021vak}. Its two-jet decays proceed through an $s$-channel vector: in the SM through the photon, $\Upsilon\to\gamma^*\to q\bar q$, with $B(\Upsilon\to\gamma^*\to q\bar q)/B(\Upsilon\to\mu^+\mu^-)=3\sum_{q=u,d,s,c}Q_q^2=10/3$ up to QCD corrections, and in our model also through $X$. We here calculate the constraint on our model parameter space from this process.

Since leptons are neutral under $X$, the dimuon rate is unaffected and the ratio of two-jet to dimuon rates measures the $X$ exchange directly. Adding the photon and $X$ amplitudes, $\mathcal A_q\propto e^2Q_bQ_q/m_\Upsilon^2+g_X^2x_bx_q/(m_\Upsilon^2-m_X^2)$, the shift of this ratio is
\begin{equation}
    \Delta R_\Upsilon=3\!\!\sum_{q=u,d,s,c}\!\Big[\Big(Q_q+3\,\frac{\alpha_X}{\alpha}\,x_bx_q\,r\Big)^2-Q_q^2\Big]
    =-24\,\frac{\alpha_X}{\alpha}\,r+432\,\Big(\frac{\alpha_X}{\alpha}\Big)^{\!2}r^2\,,
    \label{eq:DeltaR-Upsilon}
\end{equation}

\clearpage
\enlargethispage{1.5cm}

\begin{figure}[H]
    \centering
    \vspace*{-1.2cm}

    \makebox[\textwidth][c]{%
    \begin{minipage}{1.04\textwidth}
        \centering

        \begin{subfigure}[t]{\textwidth}
            \centering
            \includegraphics[width=0.74\linewidth]
            { 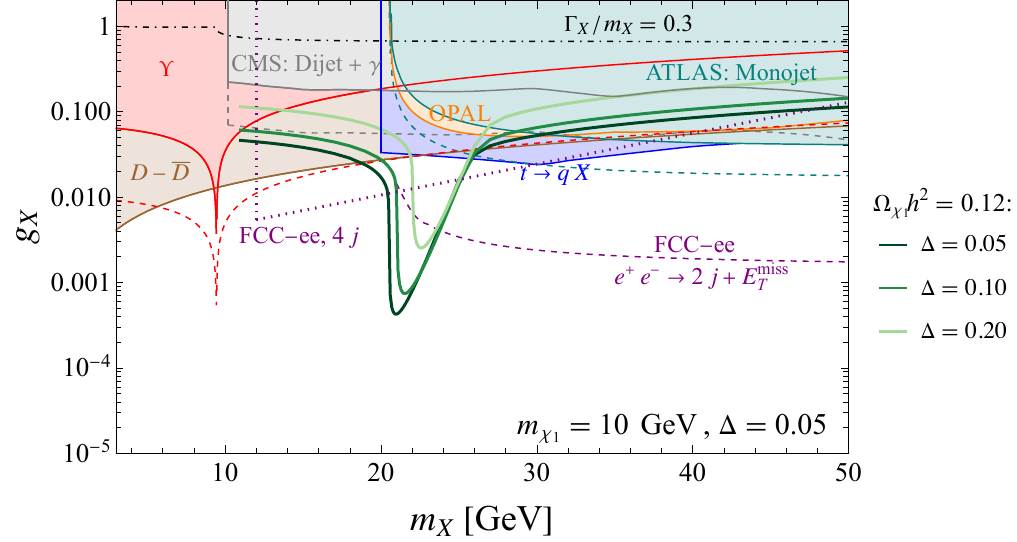}
            \caption{}
            \label{fig:gvsmDeltaHL}
        \end{subfigure}
        \vspace{0.1cm}

        \begin{subfigure}[t]{\textwidth}
            \centering
            \includegraphics[width=0.77\linewidth]
            { 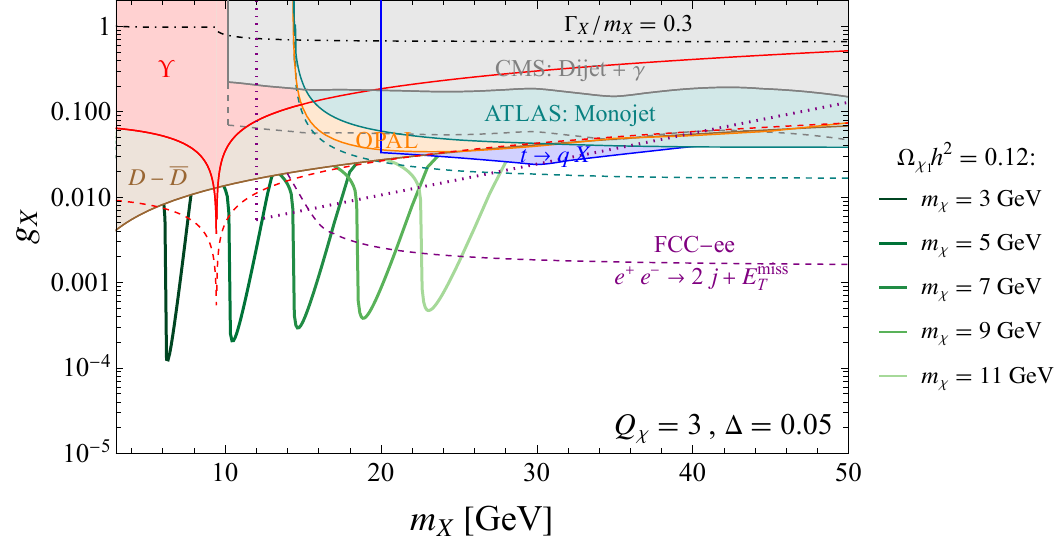}
            \caption{}
            \label{fig:gvsmmchiHL}
        \end{subfigure}
        \vspace{0.1cm}

        \begin{subfigure}[t]{\textwidth}
            \centering
            \includegraphics[width=0.74\linewidth]
            { 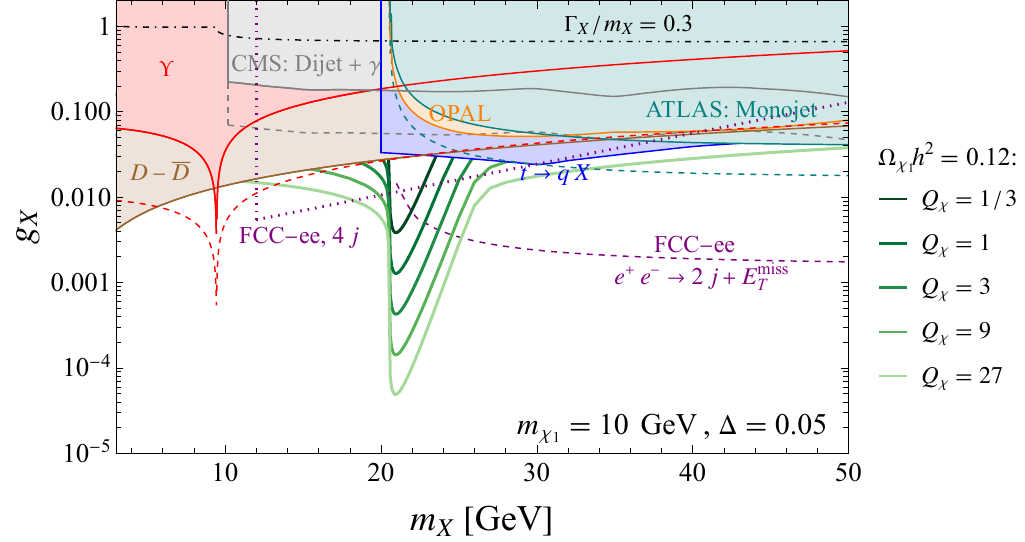}
            \caption{}
            \label{fig:gvsmQchiHL}
        \end{subfigure}

        \vspace{-0.15cm}

        \caption{Constraints and projected sensitivities at 95\% C.L. in the
        $(m_X,g_X)$ plane from exotic top decays (blue), the CMS
        dijet-plus-photon search (gray), the ATLAS monojet search
        (cyan), the OPAL search for hadronic $Z$ decays with
        missing energy (orange), $\Upsilon$ decays (red),
        $D^0$--$\bar D^0$ mixing (brown), and projected searches at FCC-ee (purple). The green
        curves indicate parameter points reproducing the observed
        relic abundance, $\Omega_{\chi_1}h^2=0.12$, for representative
        variations of (a) the mass splitting, (b) the dark-matter
        mass, and (c) the dark charge. Rare $B$ decays exclude $m_X\lesssim4.8$~GeV (Sec.~\ref{sec:additional}). Solid lines denote current
        constraints, while dashed lines denote HL-LHC, Belle II, and FCC-ee
        projections. The dash-dotted black line marks
        $\Gamma_X/m_X=0.3$. In figure (b), we used $m_{\chi_1}=7$ GeV benchmark for deriving the limits.}
        \label{fig:parameter-scans}

    \end{minipage}%
    }

    \vspace*{-0.5cm}
\end{figure}

\clearpage

\noindent where $r\equiv m_\Upsilon^2/(m_X^2-m_\Upsilon^2)$, $\alpha_X=g_X^2/4\pi$, $\alpha=\alpha(m_\Upsilon)\simeq1/132$, and the second form holds for our charges $x_q=1$, $x_b=-2$.

The sign of the linear (interference) term is fixed by $x_b\sum_qQ_qx_q=-4/3$; for a universal baryonic $Z'$ it is positive. The $\Upsilon$ wave function cancels in the ratio and the $\mathcal O(\alpha_s)$ correction to the $q\bar q$ final state is common to the photon and $X$ terms, so Eq.~\eqref{eq:DeltaR-Upsilon} is essentially free of hadronic uncertainties; finite quark masses, $Z$ exchange and kinetic mixing change the resulting bound by less than a percent~\cite{Dobrescu:2021vak}. The experimental input is the ARGUS event-shape analysis, which limits the non-electromagnetic two-jet fraction of direct $\Upsilon(1S)$ decays to below $5.3\%$ at 95\%~C.L.~\cite{ARGUS:1986nzm}; with $B(\Upsilon\to\mu^+\mu^-)=2.48\%$ this corresponds to $\Delta R_\Upsilon<2.1$~\cite{Dobrescu:2014fca}, a limit that has not been updated in four decades. Solving Eq.~\eqref{eq:DeltaR-Upsilon} for the largest allowed coupling gives
\begin{equation}
    g_X<0.099\,\sqrt{\frac{m_X^2}{m_\Upsilon^2}-1}\quad(m_X>m_\Upsilon)\,,\qquad
    g_X<0.067\,\sqrt{1-\frac{m_X^2}{m_\Upsilon^2}}\quad(m_X<m_\Upsilon)\,,
    \label{eq:Upsilon-bound}
\end{equation}
i.e.\ $g_X<0.064$, $0.036$, $0.034$, $0.077$, $0.12$ and $0.18$ at $m_X=3$, $8$, $10$, $12$, $15$ and $20$~GeV. This is a tree-level effect with a light propagator, which is why a 1986 measurement at the 5\% level competes with the LHC. Above the resonance the $X$ contribution decouples as $m_\Upsilon^2/m_X^2$ and the bound weakens linearly in $m_X$; below it the $X$ propagator becomes indistinguishable from the photon's and the bound saturates at a constant. The interference is destructive above $m_\Upsilon$ and constructive below, opposite to a universal baryonic $Z'$, so that our low-mass plateau is twice as strong as the one of Ref.~\cite{Dobrescu:2014fca}. Equation~\eqref{eq:Upsilon-bound} holds for $|m_X-m_\Upsilon|\gg\Gamma_X\simeq0.64\,g_X^2m_X$. The $\Upsilon(2S,3S)$ add weaker constraints with narrow features at $10.02$ and $10.36$~GeV~\cite{Aranda:1998fr}, and the analogous charmonium bounds at $m_X\lesssim5$~GeV, where $c$ and light quarks carry the same charge, are those of Ref.~\cite{Dobrescu:2014fca} with $g_X$ in place of the universal coupling.

If $2m_{\chi_1}<m_\Upsilon$, the same exchange gives $\Upsilon(1S)\to X^*\to\chi_1\chi_2$. Since $\chi_2$ leaves the detector before decaying throughout the thermal target (Sec.~\ref{sec:chi2-decays}), this is an invisible decay and the BaBar limit $B(\Upsilon(1S)\to{\rm invisible})<3.0\times10^{-4}$~\cite{BaBar:2009gco} applies to
\begin{equation}
    \frac{B(\Upsilon\to\chi_1\chi_2)}{B(\Upsilon\to\mu^+\mu^-)}
    =\frac{\beta_\chi^3}{4}\Big(3\,x_bQ_\chi\,\frac{\alpha_X}{\alpha}\,r\Big)^{\!2}
    =9\,Q_\chi^2\,\beta_\chi^3\Big(\frac{\alpha_X}{\alpha}\Big)^{\!2}r^2<1.2\times10^{-2}\,,
    \label{eq:Upsilon-inv}
\end{equation}
with $\beta_\chi=(1-4m_{\chi_1}^2/m_\Upsilon^2)^{1/2}$ and $\beta_\chi^3/4$ the ratio of scalar-pair to massless-fermion-pair phase space. For $m_{\chi_1}=3$~GeV and $Q_\chi=3$ this gives $g_X<0.040\,\sqrt{|m_X^2/m_\Upsilon^2-1|}$, somewhat stronger than Eq.~\eqref{eq:Upsilon-bound} and scaling as $Q_\chi^{-1/2}$.

In Fig.~\ref{fig:parameter-scans}, Eq.~\eqref{eq:Upsilon-bound} (red) is stronger than the LHC searches for $m_X\lesssim13$~GeV, where the dijet-plus-photon search has no acceptance and the monojet limit is an extrapolation, competitive with the LEP bound of Sec.~\ref{sec:LEP}, and the strongest constraint of all in the vicinity of the resonance. For $Q_\chi=3$ the thermal targets of panel (b) at these masses require $g_X\lesssim0.02$ and lie below the present bound, whereas the lower non-resonant branch of the $m_{\chi_1}=10$~GeV benchmark, $g_X\simeq0.05$, is excluded by it for $6\lesssim m_X\lesssim10.5$~GeV; both are within reach of the improvement discussed next.

\paragraph{Future projections at Belle II.}

The bound of Eq.~\eqref{eq:Upsilon-bound} rests on a 1986 measurement, and the samples to improve it exist. Belle recorded $10^8$ $\Upsilon(1S)$ decays on resonance~\cite{Belle:2025pey}, BaBar and Belle hold cleanly tagged $\Upsilon(2S,3S)\to\pi^+\pi^-\Upsilon(1S)$ samples, and dedicated $\Upsilon(nS)$ runs are part of the Belle~II programme~\cite{Belle-II:2018jsg}. The two-jet fraction of direct $\Upsilon(1S)$ decays is limited by the modelling of the $ggg$ event shape rather than by statistics; if a modern analysis controlled it at the $10^{-3}$ level, i.e.\ $|\Delta R_\Upsilon|\lesssim0.04$, the interference term of Eq.~\eqref{eq:DeltaR-Upsilon}, linear in $\alpha_X$, would become the leading sensitivity and the reach in $g_X$ would improve by a factor of five to eight, to $g_X\sim10^{-2}$ across the mass range. The dashed red line in Fig.~\ref{fig:parameter-scans} indicates this reach, obtained by rescaling the present bound by a factor of seven; it cuts through the thermal targets, including part of the resonant dips. The same tagged samples improve the invisible width, Eq.~\eqref{eq:Upsilon-inv}, which for $m_{\chi_1}<m_\Upsilon/2$ probes the dark charge directly. Both interpretations rely on $\chi_2$ leaving the detector, which Sec.~\ref{sec:chi2-decays} shows to be the case throughout the thermal target. At its upper edge, $g_X\gtrsim10^{-2}$ and $\Delta m_\chi\gtrsim0.5$~GeV, the decay $\chi_2\to\chi_1\pi^0\gamma$ instead occurs within metres, and $\Upsilon(1S)\to\chi_1\chi_2$, or continuum production through the resolved quark current, $e^+e^-\to\gamma^*\to q\bar q X^*\to$ hadrons $+\,\chi_1\chi_2$, would show up as a displaced $\pi^0\gamma$ vertex recoiling against missing energy, a signature with essentially no SM background for which Belle~II is ideally suited~\cite{Belle-II:2018jsg, Duerr:2019dmv}.

\subsubsection{$Z$ decays} 
\label{sec:LEP}

The invisible decay of the mediator can also be constrained using hadronic $Z$ decays at LEP. The OPAL collaboration analysed a sample of $4.4\times10^{6}$ hadronic $Z$ decays and searched for acoplanar\footnote{An acoplanar dijet event is one in which the two jets are not back-to-back in the plane transverse to the beam, as expected when invisible particles carry away part of the event momentum.} jets or monojet-like configurations accompanied by missing energy~\cite{OPAL:1996hvw}. The results were interpreted in terms of an invisibly decaying scalar $S_{\rm inv}$ produced through Higgs-strahlung
\begin{equation}
    e^+e^- \to Z \to S_{\rm inv} Z^\ast\,,
    \qquad
    Z^\ast \to q\bar q\,,
\end{equation}
where $Z^\ast$ denotes an off-shell $Z$ boson. In particular, Fig.~3(b) of Ref.~\cite{OPAL:1996hvw} gives the observed $95\%$ C.L. upper limit on
\begin{equation}
    R_{95}(m_S)
    =
    \frac{\sigma(e^+e^-\to S_{\rm inv}Z^\ast)}
    {
    \sigma(e^+e^-\to H^0 Z^\ast)}\,.
    \label{eq:OPAL-R95}
\end{equation}
with the denominator being the Standard-Model Higgs-strahlung cross section evaluated for (at that time) a hypothetical Higgs boson with mass $m_{H^0}=m_S$.

The UV completion of the topological portal gives the same visible final state through
\begin{equation}
    e^+e^- \to Z \to q\bar q X\,,
    \qquad
    X\to\chi_1\chi_2\,,
\end{equation}
provided that $\chi_1$ and $\chi_2$ escape the detector, which holds throughout the thermal target (Sec.~\ref{sec:chi2-decays}). We therefore obtain an approximate rate-level recast by requiring
\begin{equation}
    \sigma(e^+e^-\to q\bar qX)\,
    {\rm Br}(X\to\chi_1\chi_2)
    <
    R_{95}(m_X)\,
    \left.
    \sigma(e^+e^-\to q\bar q H^0)
    \right|_{m_{H^0}=m_X}\,.
    \label{eq:OPAL-recast}
\end{equation}
We evaluated the signal and reference cross sections with \textsc{MadGraph5\_aMC@NLO} at $\sqrt{s}=m_Z$. For the reference process, the $H^0$ mass was varied over the mediator mass range while its Standard-Model $HZZ$ coupling was kept fixed. The signal was generated at the reference coupling $g_X^{\rm ref}=0.1$, with $x_q=1$, $Q_\chi=3$, $m_{\chi_1}=1~{\rm GeV}$ and masses corresponding to the benchmark values used in Fig.~\ref{fig:parameter-scans}, summing over
the $u,d,s,c$ and $b$ final states. 

In the narrow-width approximation, the decay-chain cross section scales as $g_X^2$, such that the limits are obtained through
\begin{equation}
    g_{X,95}(m_X)
    =
    g_X^{\rm ref}
    \left[
    \frac{
    R_{95}(m_X)\,\sigma_{\rm SM}(m_X)
    }{
    \sigma_{\rm sig}(m_X,g_X^{\rm ref})
    }
    \right]^{1/2}\,.
    \label{eq:OPAL-gX}
\end{equation}
Here, $\sigma_{\rm SM}(m_X)$ denotes the Higgs-strahlung reference cross section on the right-hand side of Eq.~\eqref{eq:OPAL-recast}, while $\sigma_{\rm sig}$ includes ${\rm Br}(X\to\chi_1\chi_2)$. The solution to Eq.~\eqref{eq:OPAL-gX} results in the excluded region shown in orange in Fig.~\ref{fig:parameter-scans}, with its boundary indicated by the orange line.

\subsection{Additional signatures of the concrete model}
\label{sec:additional}

The down-aligned completion of $B_1 + B_2 - 2 B_3$ model correlates all tree-level flavour-changing couplings of
$X$.  In the convention $x_q=1$, in
which the coupling to each first- or second-generation quark is $g_X$, the
relevant entries are
\begin{equation}
  g_{tc}^L=-3g_XV_{tb}V_{cb}^*,\qquad
  g_{tu}^L=-3g_XV_{tb}V_{ub}^*,\qquad
  g_{uc}^L=-3g_XV_{ub}V_{cb}^*.
  \label{eq:flavour-couplings}
\end{equation}
Rare top decays and neutral-$D$ mixing are therefore correlated predictions
of the CKM-generating completion.

\subsubsection{Rare top decay}
Neglecting the final-state up-quark mass, the two-body decay width is
\begin{equation}
  \Gamma(t\to qX)=\frac{|g_{tq}^L|^2}{32\pi}
  \frac{m_t^3}{m_X^2}(1-r_X)^2(1+2r_X),
  \qquad r_X\equiv\frac{m_X^2}{m_t^2},\qquad q=u,c.
  \label{eq:top-fcnc-width}
\end{equation}
The $1/m_X^2$ enhancement is the longitudinal-vector contribution associated
with the non-conservation of the flavour-changing current.  Since $|V_{ub}| \ll |V_{cb}|$, the charm mode controls the present sensitivity.

ATLAS has searched for $t\to qX$, followed by $X\to b\bar b$, using
$139\,\mathrm{fb}^{-1}$ of $13\,\mathrm{TeV}$ data, and provides the observed
$95\%$ C.L. upper limit
$L_{tcbb}(m_X)=[\mathrm{Br}(t\to cX)\mathrm{Br}(X\to b\bar b)]_{95}$ for
$20\leq m_X/\mathrm{GeV}\leq160$~\cite{ATLAS:2023mcc}.  To map this limit to
our model, define $\widehat\Gamma_{tq}\equiv\Gamma(t\to qX)/g_X^2$.  Neglecting the small new contribution to the total top width, the coupling shown in
Fig.~\ref{fig:parameter-scans} is obtained from
\begin{equation}
  g_{X,95}^{(t)}(m_X)=
  \left[
    \frac{L_{tcbb}(m_X)\,\Gamma_t^{\rm SM}}
    {\mathrm{Br}(X\to b\bar b)\widehat\Gamma_{tc}}
  \right]^{1/2}.
  \label{eq:top-atlas-recast}
\end{equation}
We use $m_t=172.57\,\mathrm{GeV}$, $\Gamma_t^{\rm SM}=1.32\,\mathrm{GeV}$, and ${\rm Br}(X\to b\bar b)\simeq0.5$, which follows from the widths of Sec.~\ref{sec:freezeout} since the four-fold enhanced $b$-quark coupling balances the four light flavours; the limit scales as ${\rm Br}(X\to b\bar b)^{-1/2}$.

This curve is an approximate reinterpretation, rather than a direct ATLAS
limit on the present model. The published signal is a scalar, whereas our
mediator is a vector with different polarization and angular distributions. However, since the decay is dominated by the longitudinal polarisation, we expect small corrections to \eqref{eq:top-atlas-recast}. 

\subsubsection{Neutral-meson mixing and flavour alignment}
When both $U$ and $D$ vector-like quarks are present, tree-level $X$ exchange generates the same left-left operator in each neutral
meson system,
\begin{equation}
  {\cal H}_{\rm eff}^{\Delta F=2}=C_1^{ij}(\mu)Q_1^{ij}(\mu)+\mathrm{h.c.},
  \qquad
  Q_1^{ij}=(\bar q_i\gamma_\mu P_Lq_j)(\bar q_i\gamma^\mu P_Lq_j),
  \qquad
  C_1^{ij}(m_X)=\frac{(g_{ij}^L)^2}{2m_X^2}.
  \label{eq:deltaF2}
\end{equation}
The complex square in the Wilson coefficient retains the new weak phase.  In
our $P_L$ convention the matrix element and the corresponding amplitude are
\begin{equation}
  \langle P^0|Q_1^{ij}(\mu)|\bar P^0\rangle
  =\frac{2}{3}m_P^2f_P^2B_1^P(\mu),
  \qquad
  2|M_{12}^{X,P}|=\frac{\eta_Pm_Pf_P^2B_1^P}{3}
  \left|\frac{g_{ij}^L}{m_X}\right|^2,
  \label{eq:meson-mixing-master}
\end{equation}
where $\eta_P$ evolves $C_1^{ij}$ from $m_X$ to the scale of the lattice
matrix element.  For charm we use leading-logarithmic running to
$3\,\mathrm{GeV}$,
\begin{equation}
  \eta_D(m_X)=
  \left[\frac{\alpha_s(m_X)}{\alpha_s(m_b)}\right]^{6/23}
  \left[\frac{\alpha_s(m_b)}{\alpha_s(3\,\mathrm{GeV})}\right]^{6/25},
  \qquad
  B_1^{D,\overline{\rm MS}}(3\,\mathrm{GeV})=0.757,
  \label{eq:deltaC2-running}
\end{equation}
with the obvious single-factor expression below the bottom threshold.  We use
the lattice inputs collected in Refs.~\cite{Carrasco:2015pra,FLAG:2024cha}
and the meson masses, lifetimes, and oscillation frequencies from the
PDG~\cite{ParticleDataGroup:2022pth}.

The flavour allocation is fixed by a single complex unit vector $b$ through
Eq.~\eqref{eq:flavour-alignment-vectors}.  Explicitly,
\begin{equation}
  \begin{aligned}
  g_{uc}^L&=-3g_Xa_ua_c^*, &
  g_{ds}^L&=-3g_Xb_db_s^*,\\
  g_{db}^L&=-3g_Xb_db_b^*, &
  g_{sb}^L&=-3g_Xb_sb_b^*, &\qquad a&=V_{\rm CKM}b.
  \end{aligned}
  \label{eq:all-meson-couplings}
\end{equation}
We parameterize
$b=(z_d,z_s,1)^T/\sqrt{1+|z_d|^2+|z_s|^2}$ and maximize
$\kappa\equiv g_X/m_X$ over the four real components of $z_d$ and $z_s$.
For every trial point we evaluate the complex amplitudes in
Eq.~\eqref{eq:meson-mixing-master}.  The $D$ constraints use the HFLAV
profiles for the dispersive magnitude $x_{12}$ and phase
$\phi_2^M$~\cite{Kagan:2020vri,HFLAV:2024ctg}; the $K$ constraints use both
$\Delta m_K$ and $\epsilon_K$; and the $B_d$ and $B_s$ constraints use the
phase-independent $95\%$ envelope
$|M_{12}^{\rm NP}/M_{12}^{\rm SM}|<0.25$ from the UTfit new-physics
analysis~\cite{UTfit:2025NP}.  This last condition applies to arbitrary new
weak phases and is therefore conservative compared with using the full
two-dimensional likelihoods.

It is instructive first to restrict to the down-aligned benchmark, $b=e_3$.
Discarding phase information and requiring the new contribution not to exceed
the HFLAV upper range for $x_{12}$ gives
\begin{equation}
  g_X<0.027\qquad (m_X=20\,\mathrm{GeV}),
  \label{eq:Dmixing-conservative-number}
\end{equation}
with an approximately linear scaling with $m_X$, corrected mildly by QCD
running.  In the same benchmark the calculable new amplitude has the large
phase $\theta_X=-2\gamma+\mathcal O(\lambda^4)$ relative to the dominant charm
amplitude~\cite{Betti:2024global}.  Neglecting an accidental
cancellation with the Standard-Model CP-odd amplitude, the HFLAV phase interval
then strengthens Eq.~\eqref{eq:Dmixing-conservative-number} to
$g_X<0.0049$.

Allowing both vector-like quarks changes this conclusion without eliminating
flavour violation.  Complex down-sector rotations can move the $D$-mixing
amplitude towards the CP-even axis, while simultaneously generating $B_d$ and
$B_s$ mixing.  Maximizing over $b$ with all the constraints above gives
\begin{equation}
  g_X<0.014\qquad (m_X=20\,\mathrm{GeV}),
  \label{eq:flavour-optimized-number}
\end{equation}
at a point where the CP-odd $D$ amplitude and the $B_d$ and $B_s$ bounds are
simultaneously saturated; $K$ mixing is subleading. Thus the very strong $0.0049$ result is
a feature of strict {down} alignment, whereas the optimized mixed completion
weakens it by almost a factor of three.

The explicit vector-like-quark completion also generates flavour-changing $Z$ couplings at tree level and the same left--left meson-mixing operator through electroweak boxes with vector-like quarks at one loop~\cite{Ishiwata:2015cga}.  The contributions controlled by the same rank-one flavour spurion have no independent weak phase and add rather than cancel.  Crucially, the optimized $B_s$
bound above is saturated by $X$ exchange: rotating the non-universal $X$ charge directly generates an $X\bar sb$ coupling, whereas the generation-universal SM $Z$ charge remains diagonal under the same rotation.  A $Z\bar sb$ coupling
requires in addition the electroweak singlet admixture $\lambda^d_av_{\rm EW}/M_D$.  At fixed $b_s$, the CKM mass insertion fixes only $(\lambda^d_a/M_D)y_Dv_F$, where $v_F=v_D$ (or $v_S$ in the separated flavour sector), so this admixture can be reduced by increasing $y_Dv_F$.  At the CP-optimized point,
$|c_{10}^Z/c_{10}^{\rm SM}|\simeq2.4\times10^{-3}
(1\,\mathrm{GeV}/|y_D|v_F)^2$; hence $B_s\to\mu^+\mu^-$ is already negligible for $|y_D|v_F\gtrsim1\,\mathrm{GeV}$ and is not a problem for the optimized
solution.  We assume this perturbative hierarchy, which also keeps the completion-dependent $\Delta F=2$ terms subleading, in quoting the limits above.

For simplicity, Fig.~\ref{fig:parameter-scans} displays the conservative
magnitude-only curve of Eq.~\eqref{eq:Dmixing-conservative-number}.  This
choice does not assume a particular distribution of CKM phases between the two
Yukawa sectors.  The CP-sensitive value obtained after including the down-type
vector-like quark and optimizing the alignment, $0.014$, is only a factor of
about two below the displayed value, $0.027$, at
$m_X=20\,\mathrm{GeV}$. The local-operator treatment is used only for
$m_X\gtrsim5\,\mathrm{GeV}$; below this scale a non-local hadronic analysis is
required. 

\subsubsection{Rare $B$ decays}

For $m_X<m_B-m_K\simeq4.8$~GeV the strongest flavour constraint is not mixing but $b\to sX$. The operator of Eq.~\eqref{eq:flavour-effective-operator} that generates $\delta^u_c\simeq V_{cb}m_t$ is proportional to $\phi_2$, so the longitudinal component of $X$ couples to $\bar c_Lt_R$ like a flavon, with strength $\delta^u_c/v_D=3g_XV_{cb}\,m_t/m_X$; this is the origin of the $1/m_X^2$ enhancement in Eq.~\eqref{eq:top-fcnc-width}. A top-Yukawa loop transmits it to the down sector, giving $C_V\,X_\mu\bar s\gamma^\mu P_Lb$ with $C_V\simeq3g_XV_{cb}\,(m_t^2/8\pi^2v_{\rm EW}^2)\ln(M_U/m_t)\simeq10^{-3}g_X$ at leading logarithm, and hence, from Eq.~\eqref{eq:top-fcnc-width} with $t\to b$, $B(b\to sX)\sim20\,(g_X/10^{-2})^2(3~{\rm GeV}/m_X)^2$. The $B$ lifetimes and the exclusive $B\to K+{\rm hadrons}$ modes then require $g_X\lesssim10^{-4}$, which excludes the region $m_X\lesssim4.8$~GeV of Fig.~\ref{fig:parameter-scans} for any flavour alignment (with the $D$-type vector-like quark the coupling arises already at tree level). Above the $B$-meson mass the constraint disappears: $B\to K^{(*)}\chi_1\chi_2$ is kinematically closed for $m_{\chi_1}>2.4$~GeV, the induced $b\to sq\bar q$ operators are suppressed by $g_X^2/m_X^2$ relative to the SM, and the loop-induced contribution to $B_s$ mixing lies below the $D$-mixing bound. The viable parameter space is thus confined to $m_X\gtrsim5$~GeV.

\subsection{Prospects at FCC-ee}
\label{sec:lepton-colliders}

We have already seen, in our discussion of bottomonium decays (\S \ref{sec:upsilon}) and LEP constraints (\S \ref{sec:LEP}), how the clean collision environment afforded by a lepton collider offers complementary routes to testing this dark matter mechanism, particularly in the low-mass region. 

At a Tera-$Z$ factory such as FCC-ee, the $Z^\prime$ mediator is radiated in hadronic $Z$ decays, and can be searched for in two complementary ways: firstly, through its visible decay, as a dijet resonance inside a four-jet event (Sec.~\ref{sec:FCCee-vis}); and secondly,
through its invisible decay, as a peak in the mass recoiling against the hadronic system (Sec.~\ref{sec:FCCee-inv}). 
The latter is the more sensitive where $X\to\chi_1\chi_2$ is open; the former is far less sensitive where both apply, but it is the only one that covers the region of $m_X$ masses below the $\chi_1 \chi_2$ threshold.  

\subsubsection{Visible decays: dijet resonance}
\label{sec:FCCee-vis}

We first discuss the visible decays of the $Z^\prime$ boson, which (in all cases) is radiated from a hadronic $Z$ decay, to jets. The final state is a 4-jet event, where a pair of those jets reconstructs to the $Z^\prime$ resonance. This search is important in the region  $m_X<m_{\chi_1}+m_{\chi_2}$ where the invisible decay switches off (and the region just above this, where the invisible decay is $\beta_\chi^3$-suppressed). 

Specifically, we consider the process
\begin{equation}
    e^+e^- \to Z \to q\bar q X\,,\qquad X\to b \bar{b}\,,
    \label{eq:FCCee-signal_visible}
\end{equation}
For massless quarks with vector coupling $g_Xx_q$ the tree-level emission probability is
\begin{equation}
    \frac{\Gamma(Z\to q\bar qX)}{\Gamma(Z\to q\bar q)}=g_X^2x_q^2\,P(m_X)
    \,,\qquad
    \label{eq:FCCee-emission}
\end{equation}
where 
\begin{equation}
    P(m_X)\simeq\frac{1}{8\pi^2}\Big[\ln^2\frac{m_Z^2}{m_X^2}-3\ln\frac{m_Z^2}{m_X^2}\Big]\quad(m_X\ll m_Z)\,,
    \label{eq:FCCee-emission-2}
\end{equation}
in the limit $m_X \ll m_Z$, which is a reasonable approximation for the region where this search is most important, namely for low-mass $X$.
This ratio of probabilities is
independent of the vector or axial nature of the $Z$ coupling. We also note that  
the exact evaluation of $P(m_X)$ falls faster than the logarithm as the phase space closes. 
 Weighting by the $Z$ partial widths and the charges,  a fraction $1.65\,g_X^2P(m_X)$ of hadronic $Z$ decays radiates an $X$. This translates to $4\times10^7$ events at Tera-$Z$ for $g_X=10^{-2}$ and $m_X=15$~GeV. For our $B_1+B_2-2B_3$ concrete UV completion, half of those events have $X\to b\bar b$, where of course this channel closes below the threshold $2m_B=10.6$~GeV.

We estimate the reach at FCC-ee with a parton-shower simulation of the signal and background rates~\cite{Bierlich:2022pfr}. 
(While a parton shower simulation is good enough for this visible decay channel, it will not be good enough for the invisible channel we consider in \S \ref{sec:FCCee-inv}.)
We cluster the visible final state into exactly four Durham jets with $y_{34}>0.005$~\cite{Cacciari:2011ma}, we require two $b$-tagged jets (assuming benchmark tagging parameters $\epsilon_b=0.8$, $\epsilon_c=0.02$, $\epsilon_{uds}=0.002$~\cite{Bedeschi:2022rnj}), and we assume a jet energy resolution of $30\%/\sqrt{E}\oplus2\%$ followed by a four-momentum-constrained rescaling of the jet energies. 
We find the reconstructed $X$ resonance has a width of $7$--$8\%$, which is mostly set by the jet assignment and by the neutrinos coming from the semileptonic $b$ decays, rather than by the calorimeter resolution. 
Below $20$~GeV, the background is dominated by gluon splitting to $b\bar b$, which occurs with rate $2.5\times10^{-3}$ per hadronic decay~\cite{ALEPH:2005ab} and with $m_{b\bar b}$ falling from threshold, with a $20\%$ charm-mistag component. Above $25$~GeV, the background is dominated by primary $b\bar bgg$ events, whose $m_{b\bar b}$ peaks near $60$~GeV. 

To give an idea, we find
the signal-to-background ratio for $g_X=10^{-2}$ and $m_X=15$~GeV is $3\times10^{-3}$ at $15$~GeV, dropping to $2\times10^{-3}$ at $20$~GeV and then further to $4\times10^{-4}$ at $30$~GeV: the search is limited by the modelling of the smooth $m_{b\bar b}$ spectrum rather than by statistics, which alone would allow $g_X\sim10^{-3}$. For a residual systematic uncertainty $f$ on the background in the window, the $95\%$~C.L. reach is well described by
\begin{equation}
    g_X\lesssim1.2\times10^{-2}\,\exp\!\left[\frac{m_X-15~{\rm GeV}}{12~{\rm GeV}}\right]\left(\frac{f}{0.3\%}\right)^{1/2},\qquad 12~{\rm GeV}\lesssim m_X\lesssim50~{\rm GeV}\,,
    \label{eq:FCCee-vis-reach}
\end{equation}
A reasonable value is $f=0.3\%$, which we use to plot the dotted purple contour (labelled `FCC-ee, $4j$') in Fig.~\ref{fig:parameter-scans}. 

Unlike the missing-energy searches that we will discuss next, this reach is independent of $m_{\chi_1}$ and $Q_\chi$ up to variations in the small invisible branching ratio. The same curve applies to all panels in Fig.~\ref{fig:parameter-scans}. Between $2m_B$ and $12$~GeV, where $X\to b\bar b$ is closed or too close to threshold, only an untagged dijet search on the full four-jet combinatorial background remains, with a reach of $g_X\sim0.1$. The visible channel is far less sensitive than the recoil-mass search where $X\to\chi_1\chi_2$ is open (as we will demonstrate next), but it is the only one below the $\chi_1\chi_2$ threshold, where it covers the lower non-resonant thermal branch of the $m_{\chi_1}=10$~GeV benchmark. 
For comparison, we note that the same analysis on the $1.7\times10^7$ hadronic $Z$ decays recorded at LEP, where it would be statistics-limited, would reach $g_X\simeq0.03$ for $m_X\lesssim20$~GeV; it has not been performed.

\subsubsection{Invisible decays: recoil-mass search}
\label{sec:FCCee-inv}

We now turn to invisible decays, via the process:
\begin{equation}
    e^+e^- \to Z \to q\bar q X\,,\qquad X\to\chi_1\chi_2\,,
    \label{eq:FCCee-signal}
\end{equation}
The production of dark matter particles from the $Z^\prime$ leaves a hadronic dijet system recoiling against missing momentum (noting that $\chi_2$ escapes the detector throughout the thermal target -- see Sec.~\ref{sec:chi2-decays}). 
As is crucial to such missing energy searches, the recoil mass $m_{\rm rec}^2=(p_{e^+}+p_{e^-}-p_{\rm vis})^2$ peaks at $m_X$ which allows us to gain sensitivity. 
Two model features bound the expected sensitivity from the outset: first is simply the emission probability, still given by Eq.~\eqref{eq:FCCee-emission} above, which falls as we enter the higher $X$ mass regime (from $0.03$ at $m_X=20$~GeV to $10^{-3}$ at $50$~GeV, say). Second is the invisible branching ratio, $\Gamma(X\to\chi_1\chi_2)/\Gamma(X\to{\rm hadrons})=Q_\chi^2\beta_\chi^3/96$, which is at most $8.6\%$ for $Q_\chi=3$ and which vanishes below $m_{\chi_1}+m_{\chi_2}$ as discussed above (while being suppressed as $\beta_\chi^3$ in the resonant dip regions).

We estimate the reach this time by doing a hadron-level simulation; for channels with missing energy such as this one, using only a parton-level simulation is expected to introduce significant mismodelling. 
Thus, we generate the signal using the exact matrix element and showered and hadronised with \textsc{Pythia}~8~\cite{Bierlich:2022pfr}. The background consists of all hadronic and $\tau^+\tau^-$ decays of the $Z$ boson, including initial-state radiation ($5\times10^6$ events, supplemented by dedicated samples in which both heavy hadrons decay semileptonically, and of the two-photon process $\gamma\gamma\to q\bar q$). Every particle is passed through a fast detector response modelled on the FCC-ee concepts~\cite{Bacchetta:2019fmz}: specifically, we assume an acceptance $|\cos\theta|<0.99$, and we reconstruct charged tracks with $\sigma_p/p=0.2\%$, photons with $15\%/\sqrt{E}\oplus1\%$ and neutral hadrons with $45\%/\sqrt{E}\oplus3\%$. 

Three main findings shape the analysis. First, the background is not in fact the irreducible $q\bar q\nu\bar\nu$ final state as might be anticipated, which contributes $\sim10^6$ events in total, but instead it is neutrinos coming from semileptonic $b$ and $c$ decays. With a loose selection ($\geq7$ charged particles, $E_{\rm vis}>10$~GeV, and $|\cos\theta_{\rm miss}|<0.9$), this background gives of order $10^{10}$ to $5\times10^8$ events per 6~GeV window between $m_{\rm rec}=20$ and $50$~GeV. More than $80\%$ of them come from $Z\to b\bar b$ with two semileptonic decays, for which $m_{\rm rec}^2\simeq4E_{\nu_1}E_{\nu_2}$. Second, the detector resolution does not turn out to be an important effect, numerically: replacing smeared by unsmeared momenta changes these numbers by less than $3\%$, because the recoil-mass tails are made of genuine neutrinos and of energy lost beyond the acceptance (not of calorimeter fluctuations). Third, the neutrino background is reducible. Semileptonic decays come with a lepton and their neutrinos sit inside a jet: for two of them, these nearly balance in the transverse plane, whereas $X$ carries tens of GeV of transverse momentum and is emitted at large angle from the jets.

We introduce further cuts to reduce the backgrounds.
We veto events with an identified electron or muon above $2$~GeV ($90\%$ efficiency), and we require the missing momentum to be more than $0.5$~rad from either jet and $p_T^{\rm miss}>10$~GeV. This reduces the background by two to three orders of magnitude, while keeping $33$--$43\%$ of the signal in the mass window $[m_X-1,m_X+5]$~GeV; the same cuts remove the two-photon and $\tau^+\tau^-$ backgrounds entirely. A tight veto on $b$- and $c$-tagged events~\cite{Bedeschi:2022rnj} lowers the background by a further factor $30$--$70$ at the price of $60\%$ of the signal, which is radiated predominantly from $b$ quarks.

We find
the resulting $95\%$~C.L. reach is well modelled by the following constraint on a combination of parameters:
\begin{equation}
    g_X^2\,{\rm Br}(X\to\chi_1\chi_2)\lesssim2\times10^{-7}\qquad(20\lesssim m_X\lesssim50~{\rm GeV})\,,
    \label{eq:FCCee-inv-reach}
\end{equation}
statistics-limited and nearly flat in $m_X$ because the falling emission rate is compensated by the falling background; a $0.3\%$ background systematic weakens the reach in $g_X$ by a factor $3$--$5$ at $20$--$30$~GeV and by less than a factor two above $40$~GeV. For the benchmark of Fig.~\ref{fig:parameter-scans}(a,c) this translates into $g_X\lesssim3.5\times10^{-3},\ 2.6\times10^{-3},\ 1.9\times10^{-3},\ 1.9\times10^{-3}$ at $m_X=25,\,30,\,40,\,50$~GeV, shown as the dashed purple line, and nothing below $20.5$~GeV; the flavour veto would improve these by up to a factor two, down to $1.0\times10^{-3}$ at $40$--$50$~GeV. The remaining uncertainty is instrumental: the optimised selections retain $10^{-8}$--$10^{-7}$ of hadronic $Z$ decays, a rejection that a parametric simulation cannot vouch for and that has to be established with a full detector simulation; OPAL reached $10^{-6}$ in its acoplanar-jet search~\cite{OPAL:1996hvw}, and a residual instrumental floor at that level would degrade the reach above $40$~GeV to $g_X\simeq2\times10^{-3}$ while leaving it unchanged below $30$~GeV.

\section{Conclusions}
\label{sec:conclusions}

In Ref.~\cite{Davighi:2024zip}, we proposed a topological portal between the
QCD chiral Lagrangian and two dark pNGBs $\chi_{1,2}$ parametrising the coset $SU(2)_D/U(1)_D \cong S^2$, and showed that it provides a new way to realise GeV-scale thermal dark matter. The topological nature of the interaction offers a clean route to decoupling all indirect and direct detection constraints, because it necessarily features only the inelastic channel $\chi_1 \chi_2 \to \text{SM}$; if a large enough mass splitting is opened up, one can completely decouple both effects. We consider dark matter masses of order a few GeV, and mass splitting around the pion mass, for which this is resoundingly the case.

That analysis was formulated entirely in
the low-energy theory.  The main purpose of the present work has been to follow
the portal above the QCD scale, construct a consistent ultraviolet completion,
and determine the collider phenomenology that accompanies it.  This step is
necessary whenever the momentum transfer resolves quarks, gluons, or the
mediator itself, while the original topological interaction remains the
appropriate description for sufficiently soft processes.

In Sec.~\ref{sec:uv-general}, we identified the essential features of any such
completion, which are fixed by matching symmetry structures. The form of the coupling to QCD at high energies is fixed to be the baryon number current
that matches onto the topological current of the chiral Lagrangian. The coupling to dark matter constituents is fixed by the connection induced on the pNGB manifold.  A massive abelian
gauge field coupled to these two currents then reproduces the topological
portal when it is integrated out.  In Sec.~\ref{sec:UV-concrete}, we embedded
this construction into the SM by gauging the anomaly-free,
leptophobic combination $B_1+B_2-2B_3$ and realizing the dark pNGBs with a
weakly coupled scalar doublet, whose non-zero VEV breaks $SU(2)_D$ to $U(1)_D$.  An exact
dark parity guarantees the stability of $\chi_1$, while controlled explicit
breaking produces the small mass splitting between $\chi_1$ and $\chi_2$.  We also
gave a vector-like-quark realization of the required CKM structure and, in order to be fully general, we discussed the kinetic-mixing and Higgs portals that are allowed by the same symmetries. The radiatively generated kinetic mixing is harmless for freeze-out and production, but it opens $\chi_2\to\chi_1\ell^+\ell^-$ in competition with the topological decay $\chi_2\to\chi_1\pi^0\gamma$; in Sec.~\ref{sec:chi2-decays} we showed that both widths scale as $g_X^4/m_X^4$, that the topological mode dominates for $\Delta m_\chi\gtrsim200$~MeV, and that $\chi_2$ decays well before nucleosynthesis constraints apply yet is stable on detector scales throughout the thermal target.

This construction makes the separation between robust and model-specific
phenomenology explicit.  The coupling of the light-quark baryon current to the
off-diagonal dark current, and its matching onto the infrared topological
operator, are common to the class of completions described in
Sec.~\ref{sec:uv-general}.  The choice of $B_1+B_2-2B_3$, the {down}-aligned
generation of the CKM matrix, and the relation between the visible and dark
charges are instead properties of the concrete model in
Sec.~\ref{sec:UV-concrete}.  The latter assumptions lead to additional
predictions, including flavour-changing top couplings and neutral-$D$ mixing,
which we studied separately in Sec.~\ref{sec:additional}. Throughout the paper, we are careful to conceptually separate which predictions are necessarily tied to landing on the topological portal, and which predictions are contingent on our particular concrete realisation.

Armed with the ultraviolet completion, in Sec.~\ref{sec:freezeout} we revisited
the thermal relic abundance.  Freeze-out takes place at centre-of-mass energies
of order $m_{\chi_1}+m_{\chi_2}$, where a local interaction written only in
terms of QCD pions is generally not reliable (given we consider masses of a few GeV).  We therefore computed
$\chi_1\chi_2\to X^*\to q\bar q$ using quark degrees of freedom and the full
$X$ propagator.  We treated both non-resonant freeze-out and the resonant region
$m_X\simeq m_{\chi_1}+m_{\chi_2}$, where the velocity expansion breaks down.
The resulting thermal targets display explicitly their dependence on the
mediator mass, the dark-matter mass and splitting, and the relative dark
charge $Q_\chi$.

In Sec.~\ref{sec:LHC}, we confronted these thermal targets with the
unavoidable hadron-collider signatures predicted by the completion, that follow from its tree-level couplings to quarks and dark matter.  Visible mediator
decays to $q\bar{q}$ are best tested in the mass region  $m_X \lesssim 50 \text{~GeV}$ of interest by the CMS boosted-dijet-plus-photon search~\cite{CMS:2019xai}, while
$X\to\chi_1\chi_2$ gives a monojet and missing-momentum signature, for which we
performed an approximate recast of the ATLAS analysis~\cite{ATLAS:2021kxv}.  The two searches are
complementary: increasing the invisible branching fraction weakens the dijet
bound while strengthening the monojet signal.  The non-resonant thermal
branches, which require larger couplings, are already probed over part of the
parameter space: the branch above the coannihilation threshold by the missing-energy and top-decay searches, the one below it, where the mediator cannot decay invisibly, only by $\Upsilon(1S)$ decays and by flavour, whereas resonant coannihilation remains an important viable
region.  The flavour observables considered in
Sec.~\ref{sec:additional} provide complementary tests of the
specific {down}-aligned completion rather than of the topological portal itself.

Finally, in Sec.~\ref{sec:lepton-colliders} we turn to FCC-ee and study the the opportunities at this $e^+ e^-$ machine.  Indeed, the constraints from LEP searches for jets plus missing energy are already important, comparable in strength to the LHC bounds from dijet and monojet (especially for low mediator masses). Below the LHC reach, the strongest present constraint is the four-decade-old ARGUS limit on two-jet $\Upsilon(1S)$ decays, Sec.~\ref{sec:upsilon}, which the $b$-quark {charge of the $B_1+B_2-2B_3$ embedding} makes a sensitive probe; a modern measurement with the existing Belle and BaBar samples, or at Belle~II, could improve the reach in $g_X$ by almost an order of magnitude and cut through the thermal targets. The dark pair produced in $\Upsilon$ decays or in the continuum appears as missing energy, since the small momentum release in the $\chi_2$ decay returns us to the low-energy topological interaction and makes $\chi_2$ long-lived; only at the upper edge of the viable couplings and splittings does $\chi_2\to\chi_1\pi^0\gamma$ give rise to a striking displaced semi-visible signature, for which Belle~II is ideally suited.  

At a future high-statistics
$Z$ factory, Sec.~\ref{sec:lepton-colliders} showed that radiation of an invisibly
decaying $X$ in hadronic $Z$ decays can be searched for in the recoil-mass
spectrum, similar to the LEP strategy.  Below the $\chi_1\chi_2$ threshold, where the mediator decays visibly, a $b\bar b$ resonance search in four-jet events reaches $g_X\simeq0.01$--$0.02$ for $m_X\lesssim20$~GeV, enough to cover the lower thermal branch. A detector-level treatment of instrumental missing momentum will
be needed for a definitive sensitivity estimate, but our preliminary estimates clearly demonstrate that the enormous $Z$ sample
offers a compelling way to explore the LHC low-mass blind spot for this dark matter mechanism.

One central lesson of this paper is that different parts of the phenomenology require
different descriptions of the same portal.  The ultraviolet model controls
thermal freeze-out and hard collider production, while the infrared
topological interaction controls soft transitions between the dark states.
This connection turns the formal matching condition of~\cite{Davighi:2024zjp}, which is robustly fixed by the need to match generalised symmetry structures, into a coherent
experimental program: low-mass resonance and monojet searches at the LHC, precision bottomonium decays together with missing-energy and displaced searches at Belle II, and recoil-mass searches at a
future $Z$ factory probe complementary aspects of the construction.

\section*{Acknowledgments}

We are especially grateful to Nakarin Lohitsiri for his contribution to Ref.~\cite{Davighi:2024zjp}, which lays the foundation of the UV completion further developed here. We also thank Tim Cohen, Matthew McCullough and Ethan Torres for useful discussions. The work of JD was supported by the Science and Technology Facilities Council (STFC) through an Ernest Rutherford Fellowship, under grant UKRI/ST/C002428/1, and was partially supported by the STFC HEP consolidated grant ST/X000664/1. The work of AG and LS was supported by the Swiss High Energy Physics initiative for the FCC (CHEF), with funding provided specifically by SERI and by the University of Basel. The work of NS is supported by the Italian MUR through the FIS 2 project FIS-2023-01577 (DD n. 23314 10-12-2024, CUP C53C24001460001), and by Istituto Nazionale di Fisica Nucleare (INFN) through the Theoretical Astroparticle Physics (TAsP) project and the INFN Iniziativa Specifica APINE.

\bibliographystyle{JHEP}
\bibliography{references}
\end{document}